\documentclass[aps,prb,reprint,floatfix,superscriptaddress,amsmath,amssymb,longbibliography]{revtex4-2}
\usepackage{graphicx}
\usepackage{bm}
\usepackage{placeins}
\usepackage{CJK}
\usepackage[colorlinks=true,linkcolor=blue,citecolor=blue,urlcolor=blue,filecolor=blue]{hyperref}
\begin{document}
\begin{CJK}{UTF8}{ipxm}

\title{Fermi-volume anomaly in quartet condensates}

\author{Yixin Guo (郭一昕)}
\email{yixin.guo@riken.jp}
\affiliation{RIKEN Nishina Center for Accelerator-Based Science, Wako 351-0198, Japan}

\date{\today}

\begin{abstract}
We investigate the relation between density and Fermi volume in a
four-component Fermi system with quartet order. For a regular
zero-temperature branch without pair order, we express the correction
to the Fermi-volume count in terms of the response of the Luttinger-Ward
functional to a phase winding of the quartet source. Its leading term
is a weighted difference between four-particle and four-hole states.
For a Gaussian vertex and parabolic dispersion, we find that the
correction expands the Fermi volume in one dimension and contracts it
in three dimensions in the parameter range studied. The sign in two
dimensions depends on the momentum cutoff. We solve the one-dimensional
Dyson equation at fixed density and compare the counting correction
obtained from the Fermi points with that evaluated from the dressed
spectra at low temperature. At fixed quartet amplitude and peak
normalization of the vertex, the correction depends strongly on the
vertex width, whereas the fermionic entropy changes little. We also
examine the linear pairing response, which can limit the continuation
of the unpaired branch to zero temperature.
\end{abstract}

\maketitle
\end{CJK}

\section{Introduction}\label{sec:intro}

Quartet condensation establishes coherence among four fermions. In
this work, we consider quartet order in the absence of pair
condensation. Quartet states have been studied in
condensed-matter~\cite{BergFradkinKivelson2009,Ge2024,GuoEH2022},
cold-atom~\cite{Wu2005,Capponi2008,Liu2023}, and nuclear
physics~\cite{Ropke1998,Tohsaki2001,Sogo2010,GuoNM2022,GuoNZ2025}.
Recent work has constructed an exactly solvable model with quartet order
but no pair condensate~\cite{Jin2026}, and developed a variational
description of dilute fermion quartets~\cite{XuWu2026}. Since the quartet order
parameter is quartic in the fermion fields, it couples a fermion to a
continuum of three-fermion states rather than to a single partner, so that
the Fermi surface can survive the
condensation~\cite{BergFradkinKivelson2009,Jiang2017,GnezdilovWang2022,GuoTL2026},
while other quartet states have gapped fermions~\cite{Jiang2017,Wan2026,GuoTL2026}.
When a Fermi surface survives, can its volume still count the particles?
The usual density-volume constraint~\cite{Luttinger1960,LuttingerWard1960,Oshikawa2000}
is not enforced by the frequency-shift argument in a static quartet
background, because a quartet vertex absorbs four fermions. Gnezdilov and Wang identified the resulting Luttinger-Ward
response but left the counting correction unevaluated because it depends
on microscopic details~\cite{GnezdilovWang2022}. The broken shift invariance
alone fixes neither its magnitude nor its sign.

In our previous work~\cite{GuoTL2026}, we compared the single-particle
spectra of momentum-locked and continuum quartet states. Here we use a
finite-width vertex to investigate how the continuum affects the density
and the Fermi-volume count. We consider a phase-coherent quartet branch
without pair order and calculate the correction to the usual count. Its leading term is the $1/E_4^2$-weighted imbalance between
four-particle and four-hole states [Eq.~(\ref{eq:J})]. The imbalance
identifies the continuum contribution to the difference between the
density and the surviving Fermi-volume count. For the specified parabolic-band Gaussian model,
band convexity and radial phase space compete locally as $d-2$: the
computed Fermi points move outward in one dimension and the surface
shrinks in three, whereas the two-dimensional sign depends on the cutoff
(Fig.~\ref{fig:luttinger}). We relate the correction to a winding of the
quartet source, making the known shift response
explicitly evaluable. Bare-line calculations approach the leading term
as the quartet amplitude and temperature tend to zero, and
low-temperature self-consistent spectra support its sign
and width dependence. At fixed vertex peak normalization, the count is
much more sensitive to width than the fermionic entropy, separating
continuum occupation from low-energy quasiparticle thermodynamics.
Linear pairing tests delimit the zero-temperature continuation of this
unpaired branch.

This paper is organized as follows. In Sec.~\ref{sec:model}, we introduce
the Hamiltonian and the four-line approximation. In
Sec.~\ref{sec:luttinger}, we derive the counting relation and discuss the
dimension dependence of its leading term. Section~\ref{sec:results}
presents the self-consistent solutions and the comparisons of counting,
spectral, and thermodynamic quantities. We also examine the linear
response to pair and density-wave order. Finally, we give a summary and
perspectives in Sec.~\ref{sec:summary}. In the following, we take
$\hbar=k_{\rm B}=1$.

\section{Theoretical framework}\label{sec:model}

\subsection{Hamiltonian and quartet vertex}
We consider four distinguishable fermion components $a$, $b$, $c$, and $d$
with equal masses and densities, which are coupled by the four-body
interaction $g\,a^\dagger b^\dagger c^\dagger d^\dagger dcba$~\cite{GuoTL2026}.
Here we introduce a finite-width form factor. The single-particle energy is
$\xi_k=k^2/(2m)-\mu$. The decoupling in the quartet channel gives the
coupling to a static condensate, which reads
\begin{equation}
H_\Delta=\frac1L\sum_{k_1k_2k_3}\Delta^*\phi(k_1,k_2,k_3,k_4)\,
d_{k_4}c_{k_3}b_{k_2}a_{k_1}+{\rm h.c.},
\label{eq:HD}
\end{equation}
where $L$ is the system length and $k_4=-k_1-k_2-k_3$. The form factor is
\begin{equation}
\begin{aligned}
\phi&=\exp\Bigl[-\frac{u^2+v^2}{8\kappa^2}\Bigr],\qquad
s=\frac{k_1+k_3}{2},\\
u&=\frac{k_1-k_3}{\sqrt2},\qquad v=\frac{k_2-k_4}{\sqrt2}.
\end{aligned}
\label{eq:vertex}
\end{equation}
The variables $(s,u,v)$ are the Jacobi coordinates of the two pairs $(ac)$
and $(bd)$ with $k_{1,3}=s\pm u/\sqrt2$ and $k_{2,4}=-s\pm v/\sqrt2$, and the
transformation from $(k_1,k_2,k_3)$ has a unit Jacobian. Every leg is
restricted to $|k_i|\le K$. The width $\kappa$ limits the
relative momentum within each pair but not the pair-center momentum $s$,
so that the four-particle continuum extends in $s$ up to the cutoff, which
matters in Sec.~\ref{sec:otherdim}.
We write $Q=|\Delta|^2$ and keep $\phi(0,0)=1$ throughout. In one
dimension, without the leg cutoff,
\begin{equation}
\int\frac{du\,dv}{(2\pi)^2}|\phi|^2=\frac{\kappa^2}{\pi}.
\end{equation}
At fixed $Q$ the integrated vertex weight therefore vanishes as
$\kappa\to0$. A finite-strength locked reference requires a separate
normalization or amplitude rescaling, and is not this ordinary limit.
Even with a probability-normalized vertex, the limit is nonuniform at
the bare Fermi point, where the occupation factors are discontinuous.
The discrete three-line level and hard gap of the locked model in
Ref.~\cite{GuoTL2026} are used only as a separately defined comparison.
Finite width allows continuous internal-energy support when the internal
lines can approach the Fermi points. That support alone does not imply
a zero-frequency sign change of every dressed propagator.

\subsection{Luttinger-Ward functional}
\label{sec:model-functional}
At order $Q$, the condensate contributes one skeleton diagram to the
Luttinger-Ward functional~\cite{LuttingerWard1960,Baym1962}, namely the
four-line term
\begin{equation}
\begin{aligned}
\Phi_Q&=-Q\,\Pi_4[G](0),\\
\Pi_4[G](\Omega)&=\int_W\!\!\int\frac{d^3\omega}{(2\pi)^3}\,
G_aG_bG_cG_d\Big|_{\sum_i\omega_i=\Omega}.
\end{aligned}
\label{eq:Pi4}
\end{equation}
Equation~(\ref{eq:Pi4}) is written at $T=0$, and the frequency integrals
become Matsubara sums at $T>0$. Here, $\int_W\equiv\int
ds\,du\,dv\,(2\pi)^{-3}|\phi|^2$ is taken over the cutoff domain, and the
components $a$, $b$, $c$, and $d$ are attached to $k_1$, $k_2$, $k_3$, and
$k_4$, respectively, with $G_i=G_i(k_i,i\omega_i)$. The normal propagators
contain no charge-$2e$ anomalous component. With the convention $\Sigma_i=\delta\Phi/\delta G_i$,
the frequency integrals give the self-energy of the component $a$ as
\begin{equation}
\begin{aligned}
\Sigma_a(k,z)&=Q\int\frac{du\,dv}{(2\pi)^2}\,e^{-(u^2+v^2)/(4\kappa^2)}
\frac{\mathcal N_{bcd}}{z+\xi_{k_2}+\xi_{k_3}+\xi_{k_4}},\\
\mathcal N_{bcd}&=f_2f_3f_4+(1-f_2)(1-f_3)(1-f_4),
\end{aligned}
\label{eq:sigma}
\end{equation}
for free internal lines, where $f_i=f(\xi_{k_i})$ and $k_1=k$. The
integration obeys the leg cutoff. With dressed lines, the same convolution
is taken over the spectral functions $A_i(k,\omega)=-{\rm Im}\,G_i^R(k,\omega)/\pi$. At $T=0$, only
three holes or three particles contribute, and the continuum reaches zero
energy whenever the three internal lines can sit at the Fermi points. The
grand potential is given by
\begin{equation}
\Omega=\sum_i\bigl[-{\rm Tr}\ln(-G_i^{-1})-{\rm Tr}\,\Sigma_iG_i\bigr]+QC+\Phi_Q ,
\label{eq:Omega}
\end{equation}
where $C=m\kappa/(4\sqrt\pi)-mx/(4\pi)$ is the inverse coupling. Here,
$m\kappa/(4\sqrt\pi)$ is the vacuum bubble at zero energy without the leg
cutoff, so that $x$ is the inverse zero-energy four-body scattering length
in the convention $T_4(0)=4\pi/(mx)$~\cite{GuoTL2026}, and a four-body
bound state of energy $-E_B$ in vacuum satisfies $\Pi_4^{\rm vac}(-E_B)=C$.
At a self-consistent solution,
$\Omega$ is stationary in $G$. The amplitude is fixed by
$\partial\Omega/\partial Q=C-\Pi_4[G](0)\equiv R_Q=0$, and the chemical
potential is fixed by the density.

\subsection{Parameters}
In units with $m=1/2$ and $k_{\rm F}=E_{\rm F}=1$ at the density
$n=4k_{\rm F}/\pi$, we take $K=4k_{\rm F}$, $\kappa=k_{\rm F}/2$, and
$x=0.710$, for which $C=0.0070118$. With the leg cutoff, this corresponds
to the scattering parameter $x_K=0.664$ in the same convention. The
temperatures are $T=0.05$, $0.025$, and $0.0125E_{\rm F}$.

For the Gaussian vertex without the cutoff, a vacuum bound state exists for
$0<x<\kappa\sqrt\pi=0.886k_{\rm F}$, and at $x=0.710$ it is deep, with
$E_B=4.2E_{\rm F}$ for $|k_i|\le K$ and $7.1E_{\rm F}$ without the cutoff.
The calculations below retain the four-line skeleton and follow the
unpaired branch with a Fermi sea. This branch provides the setting for
the numerical tests in Sec.~\ref{sec:results} of the counting relation
derived in Sec.~\ref{sec:luttinger}.

\section{Fermi-volume counting}\label{sec:luttinger}

\subsection{Frequency shift and quartet response}
\label{sec:luttinger-derivation}
At $T=0$, the identity $G=\partial_{i\omega}\ln G^{-1}+G\,\partial_{i\omega}\Sigma$
splits the density~\cite{Luttinger1960,LaNave2025} as
\begin{equation}
\begin{aligned}
n&=n_L+I_2,\\
n_L&=\sum_i\int\frac{d^dk}{(2\pi)^d}\frac{d\omega}{2\pi}e^{i\omega0^+}
\partial_{i\omega}\ln G_i^{-1},\\
I_2&=\sum_i\int\frac{d^dk}{(2\pi)^d}\frac{d\omega}{2\pi}\,G_i\,\partial_{i\omega}\Sigma_i.
\end{aligned}
\end{equation}
We assume a homogeneous phase-coherent background, a normal propagator
with no charge-$2e$ anomalous component, and a differentiable branch of
$\Phi[G;\Delta]$ satisfying $\Sigma_i=\delta\Phi/\delta G_i$.
Here the absence of a charge-$2e$ anomalous propagator does not imply an
unbroken particle-number $U(1)$ symmetry: quartet order breaks that
symmetry in the selected condensate background. The counting relation
is derived within this restricted branch.
The propagator is causal and canonically normalized,
$G_i(k,z)\sim1/z$, and the high-frequency boundary term in the following
integration by parts vanishes. The frequency integrals, their shift,
and the derivative of $\Phi$ must exist, or be regulated consistently
with this shift. No finite quasiparticle residue is assumed.

The definition of $n_L$ above is a frequency-winding count. For its usual
volume interpretation, additionally take $G_i^{-1}(k,0)$ to be real
away from its Fermi-surface zeros, with the canonical occupied-side
sign and no other Green-function zero surfaces. For the balanced
one-dimensional branch with one pair of Fermi points this gives
$n_L=4k^*/\pi$. If these extra conditions fail, the winding count
must be retained instead of identifying it with a single Fermi volume.

Integrating by parts in frequency, one has
$I_2=-\sum_i{\rm Tr}[\Sigma_i\partial_{i\omega}G_i]$. With
$\Sigma_i=\delta\Phi/\delta G_i$ and
$\partial_{i\omega}G=-i\,dG^\epsilon/d\epsilon$, where
$G^\epsilon(i\omega)=G(i\omega+i\epsilon)$, this reads
\begin{equation}
I_2=i\,\frac{d}{d\epsilon}\Phi[G^\epsilon]\Big|_{\epsilon=0},
\end{equation}
as noted for charge-$4e$ order in Ref.~\cite{GnezdilovWang2022}.
Number-conserving vertices are invariant under the common shift. At a
condensate vertex, the four lines carry a fixed total
frequency, and after the change of variables $\omega_j\to\omega_j-\epsilon$
the shift reappears as $4\epsilon$ in that total. We denote by
$\Phi[G;\Delta_\Omega]$ the functional with opposite frequency
injections $\pm i\Omega$ at conjugate quartet vertices, choosing the
orientation in which its four-line term is $-Q\Pi_4(\Omega)$.
The common shift changes each quartet vertex by four times the shift
and its conjugate by the opposite amount. Consequently,
$\Phi[G^\epsilon;\Delta]=\Phi[G;\Delta_{4\epsilon}]$, diagram by diagram,
and thus
\begin{equation}
n-n_L=4i\,\partial_\Omega\Phi[G;\Delta_\Omega]\big|_{\Omega=0}\equiv4n_Q ,
\label{eq:general}
\end{equation}
where the derivative is taken at fixed $G$
and $|\Delta|$. Here, $4n_Q$ is the Luttinger anomaly
$I_2$, namely the response of $\Phi$ to a uniform winding of the condensate
phase in imaginary time. It is not the density of condensed quartets,
because the gauge transformation $\psi\to e^{i\theta/4}\psi$ fixes the total
coefficient of the phase winding in the action to $n/4$, and $n_Q$ can have
either sign. When there is no Fermi sea and $n_L=0$, one has $4n_Q=n$. Since only the structure of the
functional is used, Eq.~(\ref{eq:general}) holds for the exact functional
of fermions in a static condensate and its $\Phi$-derivable skeleton
truncations under the assumptions above. Shifting two components by
$+\epsilon$ and $-\epsilon$
leaves every condensate vertex invariant, because each quartet contains
one fermion of each component. If the other vertices also conserve each component, the relative
counts obey the usual relation, with the same conditions for their
Fermi-volume interpretation~\cite{SachdevYang2006,Powell2005}. In the four-line closure, the spectral convolution can itself be evaluated
before differentiating. Writing $E=\sum_j\omega_j$ and
$D=\Theta_{pppp}-\Theta_{hhhh}$, the frequency orientation in Fig.~\ref{fig:luttinger}(a) gives
\begin{equation}
\Pi_4[G](\Omega)=\int_W\!\!\int\prod_{j=1}^4d\omega_j\,A_{k_j}(\omega_j)
\frac{D}{E-i\Omega}.
\label{eq:sourcebubble}
\end{equation}
All four positive or all four negative energies contribute at $T=0$.
With $\Phi_Q=-Q\Pi_4$, taking $-4iQ\partial_\Omega$ produces the signed
$1/E^2$ weight. The resulting correction is

\begin{equation}
n-n_L=4Q\int_W\!\!\int\prod_{i=1}^4d\omega_i\,A_{k_i}(\omega_i)\,
\frac{\Theta_{pppp}-\Theta_{hhhh}}{(\omega_1+\omega_2+\omega_3+\omega_4)^2},
\label{eq:dressed}
\end{equation}
where $\Theta_{pppp}$ ($\Theta_{hhhh}$) requires all four $\omega_i>0$
($<0$).

\subsection{Leading four-particle and four-hole weights}
With free lines, one has $A_k(\omega)=\delta(\omega-\xi_k)$, and
Eq.~(\ref{eq:dressed}) reduces to
\begin{equation}
\begin{aligned}
n-n_L&=4Q(J_{4p}-J_{4h})+O(Q^2),\\
J_{4p,4h}&=\int_W\frac{\Theta_{pppp,hhhh}}{E_4^2},\\
E_4&=\frac{4s^2+u^2+v^2}{2m}-4\mu ,
\end{aligned}
\label{eq:J}
\end{equation}
where $\Theta_{pppp}$ ($\Theta_{hhhh}$) now selects momenta outside
(inside) the free Fermi sea. We describe the integration over these
domains in Appendix~\ref{app:count}. At the parameters of the
$T=0.025E_{\rm F}$ root in Table~\ref{tab:roots}, we obtain
$J_{4p}=5.474\times10^{-4}$, $J_{4h}=1.681\times10^{-3}$, and the leading
correction $4Q(J_{4p}-J_{4h})=-0.01200$.
For fixed $\mu<0$ and sufficiently small $Q$, the free reference band and its
perturbative continuation have no Fermi sea, so that $n_L=0$ and $J_{4h}=0$.
The leading density is $n=4QJ_{4p}+O(Q^2)$.
The four-particle admixture has amplitude
$\sqrt Q\,e^{-(u^2+v^2)/(8\kappa^2)}/E_4$ and norm $QJ_{4p}$, with four
fermions contributing per unit of this weight.

\subsection{Separating boundary motion from continuum occupation}
\label{sec:countsplit}
The leading counting correction can also be derived directly from
the grand potential.
Keep $Q$, $\kappa$, $K$, and the bare dispersion $\varepsilon_k$ fixed when
differentiating with respect to $\mu$. With free lines at $T=0$, let
$D=\prod_j\Theta(\xi_j)-\prod_j\Theta(-\xi_j)$. Then
$\Pi_4^{(0)}(0)=\int_W D/E_4$ and the first-order density change is
$\delta n^{(1)}=Q\partial_\mu\Pi_4^{(0)}(0)$. Differentiating the denominator
uses $\partial_\mu E_4=-4$ and gives $4Q(J_{4p}-J_{4h})$. Differentiating
the occupation boundaries gives
\begin{equation}
\delta n^{(1)}=
-\sum_\alpha\int\frac{d^dk}{(2\pi)^d}
\delta(\xi_k)\,\Sigma_\alpha^{(1)}(k,0)
+4Q(J_{4p}-J_{4h}).
\label{eq:densitysplit}
\end{equation}
Here $\Sigma^{(1)}$ includes its factor $Q$. For example,
$\partial_\mu D=-\sum_j\delta(\xi_j)
[\prod_{\ell\ne j}\Theta(\xi_\ell)+\prod_{\ell\ne j}\Theta(-\xi_\ell)]$.
On each boundary $E_4$ is the three-line energy in
Eq.~(\ref{eq:sigma}). On the other hand, expanding the Fermi-volume count
$\sum_\alpha\int d^dk\,(2\pi)^{-d}
\Theta[-\xi_k-\Sigma_\alpha^{(1)}(k,0)]$ identifies the first term in
Eq.~(\ref{eq:densitysplit}) as $\delta n_L^{(1)}$. Hence
$\delta n^{(1)}-\delta n_L^{(1)}=4Q(J_{4p}-J_{4h})$.
In one dimension, the same boundary term is
$\delta n_L^{(1)}=-4\Sigma^{(1)}(k_\mu,0)/(\pi v_\mu)$ for four balanced,
inversion-symmetric components.

This decomposition distinguishes movement of the Fermi boundary from
changes of the continuum occupation. Adding or removing four fermions
weights the particle and hole sectors with opposite signs. Equal weighted
sectors give no leading anomaly, even with a nonzero quartet source.
The vertex charge fixes the factor four. The band, occupation constraints,
and vertex weights determine the correction. A frequency-shift identity
alone does not supply those microscopic weights.

As an independent numerical illustration at $\mu=1.005488E_{\rm F}$,
$\kappa=0.5k_{\rm F}$, and $K=4k_{\rm F}$, the density, boundary, and
continuum coefficients per unit $Q$ are $-0.001433$, $+0.003100$, and
$-0.004533$, respectively, in the one-dimensional units used here.
The first uses a chemical-potential finite difference of the bubble with
analytic $s$ integration. The second uses a separate on-shell
self-energy quadrature with the occupation boundaries resolved.
They satisfy Eq.~(\ref{eq:densitysplit}) at the displayed precision.

\subsection{Dimension, curvature, and cutoff}
\label{sec:otherdim}
\begin{figure}[!t]
\includegraphics[width=\columnwidth]{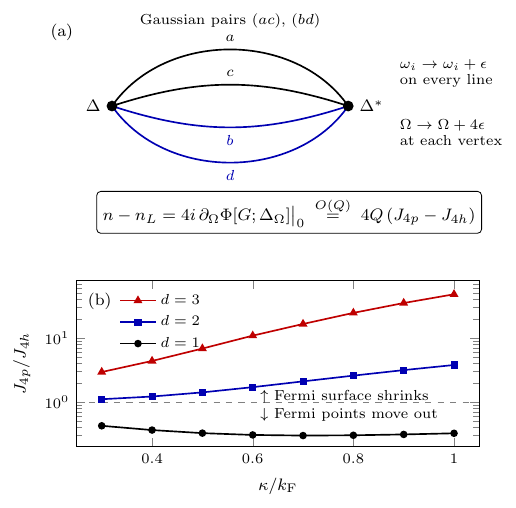}
\caption{Modified Luttinger relation and its dimension dependence.
(a)~Four-line term of the condensate [Eq.~(\ref{eq:Pi4})]. A common
frequency shift $\epsilon$ of all lines shifts the frequency at each
condensate vertex by $4\epsilon$, which gives Eq.~(\ref{eq:general}). The
first-order form [Eq.~(\ref{eq:J})] is also shown. (b)~Ratio of the
four-particle and four-hole integrals in Eq.~(\ref{eq:J}) as a
function of the quartet width $\kappa$ in one, two, and three dimensions.
At fixed density, ratios below and above unity imply a leading
expansion and contraction of the Fermi-volume count, respectively. Here, $K=4k_{\rm F}$ and $\mu=1.0055E_{\rm F}$. In two
dimensions, the ratio drops below unity for $K=2k_{\rm F}$ and
$\kappa\le0.5k_{\rm F}$ (Table~\ref{tab:cutoff}).}
\label{fig:luttinger}
\end{figure}

For the extension to $d$ dimensions, $s$, $u$, and $v$ are vectors and
$\int_W$ denotes the cutoff integral with measure
$d^ds\,d^du\,d^dv\,|\phi|^2/(2\pi)^{3d}$. We use the same form factor
and parabolic dispersion. The integration is described in
Appendix~\ref{app:count}.
For narrow quartets, $\kappa\ll k_\mu$, the four momenta lie near $\pm s$,
with $k_\mu=\sqrt\mu$ in the present units. At fixed $u$ and $v$,
the radial measure for the pair-center momentum is $|s|^{d-1}d|s|$.
For the collinear configuration $u=v=0$, we compare momenta at
$|s|=k_\mu\pm\delta$ with $\delta>0$. Band convexity favors holes
through the energy denominator, and the radial measure favors particles.
Their weight ratio is
\begin{equation}
\begin{aligned}
\frac{w_+}{w_-}&=1+2(d-2)\frac{\delta}{k_\mu}
+O\!\left(\frac{\delta^2}{k_\mu^2}\right),\\
w_\pm&=\frac{(k_\mu\pm\delta)^{d-1}}
 {\{4[(k_\mu\pm\delta)^2-k_\mu^2]\}^2}.
\end{aligned}
\end{equation}
The leading local imbalance is proportional to $(d-2)\delta/k_\mu$
and therefore vanishes in two dimensions. Figure~\ref{fig:luttinger}(b)
shows the full integrals. The ratio is
$0.30$--$0.42$ in one dimension and $2.9$--$48$ in three dimensions for
$\kappa=0.3$--$1.0k_{\rm F}$ at $K=4k_{\rm F}$. Since the vertex does not depend on $s$, $J_{4p}$ depends on
the cutoff, whereas $J_{4h}$ is cutoff independent for $K>k_\mu$. As shown in Table~\ref{tab:cutoff}, the
ratio in $d=2$ is close to unity for narrow quartets and drops below unity
at $K=2k_{\rm F}$ for $\kappa\le0.5k_{\rm F}$, namely to $0.81$--$0.97$,
whereas the ratio in $d=3$ stays above $1.3$ for
$K=2$--$8k_{\rm F}$. Therefore, in this model and parameter range the Fermi points move
outward in $d=1$ and the Fermi surface shrinks in $d=3$, whereas in $d=2$
the sign is set by the cutoff and the shape of the vertex.

\section{Results and discussion}\label{sec:results}

\subsection{Self-consistent quartet branch}\label{sec:numerics}
We solve the Dyson equation together with the amplitude and density
conditions, then test the resulting stationary states locally.

Each $G_k$ is represented by discrete spectral energies and positive
weights $(e_j,w_j)$ with $\sum_jw_j=1$ and $\sum_jw_je_j=\xi_k$, on a uniform
momentum grid of $N_k=2049$ or $4097$ points on $[-K,K]$. The three-line
convolution is evaluated on an energy grid of step $h=1/512$ (and $1/1024$
at $T=0.0125$) over the source window $[-64,64]E_{\rm F}$. The Dyson
equation is solved in the Matsubara representation with an adaptive
frequency set, while the discrete spectral energies need not lie on that grid. The inner
iteration is damped as $G\to(1-\alpha)G+\alpha G_{\rm new}$. Since the
undamped iteration cycles at $T=0.025$, it is replaced by $\alpha=0.5$
there. No artificial spectral broadening enters the iteration.

\label{sec:root-audits}
At fixed coupling and density, we follow the unpaired branch as the
temperature is successively halved from $0.05E_{\rm F}$ to
$0.0125E_{\rm F}$. For each temperature, the Dyson equation is solved
together with $R_Q=C-\Pi_4[G](0)=0$ and $n=4k_{\rm F}/\pi$ using a damped
two-dimensional Newton iteration for $(Q_*,\mu_Q)$. We recompute the kernel and use independent spectral probes to check the Dyson, amplitude, and density
residuals at each solution. The branch parameters and free-energy
differences are listed in Table~\ref{tab:roots}(a). The amplitude is an
output of these coupled equations, increasing from $Q_*=2.5069$ to
$2.6829$ on $N_k=4097$ as the temperature decreases. These solutions
provide the spectra used in the counting and entropy comparisons.

Thermodynamic response is evaluated separately by re-solving the Dyson
equation at displaced $(Q,\mu)$ on coarse and fine finite-difference
stencils. We compare $\partial\Omega/\partial Q$ and
$\partial\Omega/\partial\mu$ with $R_Q$ and $-n$, respectively, and test the
Maxwell relation $\partial n/\partial Q=\partial\Pi_4/\partial\mu$.
The fixed-$Q$ density response and the fixed-density amplitude curvature
are positive on the evaluated response stencils. We also compare the free reference density and check the Dyson equation
in the time domain.
Table~\ref{tab:roots}(b) reports the density response and relative Maxwell mismatch
for the completed stencils. At $T=0.0125$, some displaced points do not
reach the required Dyson tolerance at the lowest Matsubara frequency,
so response values are not reported for that temperature.

\begin{table}[t]
\caption{Self-consistent quartet branch and thermodynamic response.
(a)~Branch parameters and free-energy density $\Delta f$ within the
four-line approximation relative to the free normal gas at the same
density, in units of $E_{\rm F}k_{\rm F}$.
(b)~Fixed-$Q$ density response and relative Maxwell mismatch between
$\partial n/\partial Q$ and $\partial\Pi_4/\partial\mu$, normalized by the
larger absolute derivative, from the fine finite-difference stencils.}
\label{tab:roots}
\begingroup
\centering
(a) Self-consistent branch\par\smallskip
\begin{ruledtabular}
\begin{tabular}{lccc}
$T/E_{\rm F}$, $N_k$ & $Q_*$ & $\mu_Q/E_{\rm F}$ & $10^4\Delta f$\\
\hline
$0.05$, $2049$ & $2.5116$ & $1.006722$ & $-4.0629$\\
$0.05$, $4097$ & $2.5069$ & $1.006715$ & $-4.0487$\\
$0.025$, $2049$ & $2.6528$ & $1.005496$ & $-4.5420$\\
$0.025$, $4097$ & $2.6480$ & $1.005488$ & $-4.5270$\\
$0.0125$, $4097$ & $2.6829$ & $1.005185$ & $-4.6495$\\
\end{tabular}
\end{ruledtabular}
\par\medskip
(b) Thermodynamic response\par\smallskip
\begin{ruledtabular}
\begin{tabular}{lcc}
$T/E_{\rm F}$, $N_k$ & $\partial n/\partial\mu$ & Maxwell (relative)\\
\hline
$0.05$, $2049$ & $0.63666$ & $7.9\times10^{-5}$\\
$0.05$, $4097$ & $0.63666$ & $3.9\times10^{-6}$\\
$0.025$, $2049$ & $0.63566$ & $2.0\times10^{-4}$\\
$0.025$, $4097$ & $0.63569$ & $1.0\times10^{-4}$\\
\end{tabular}
\end{ruledtabular}
\endgroup
\end{table}

Refining $N_k$ from $2049$ to $4097$ at $T=0.05$ and $0.025E_{\rm F}$
lowers the root's $Q$ by $4.74\times10^{-3}$ and $4.75\times10^{-3}$,
respectively. It raises $\Delta f$ by $1.4\times10^{-6}$ and
$1.5\times10^{-6}$. The grid dependence of $\Delta f$ is
therefore approximately a temperature-independent offset over these
two temperatures, which is useful for the entropy comparison in
Sec.~\ref{sec:width-states}.

We measure low-energy spectral weight with the midpoint estimator
$D_\epsilon=\int dk\,(2\pi)^{-1}\int d\omega\,
A_k(\omega)/[2\pi\epsilon\cosh(\omega/2\epsilon)]$.
Positivity implies $D_\epsilon\le M/[2\pi\epsilon\cosh(E_g/2\epsilon)]$
for a hard gap $E_g$, where $M=\int dk\,(2\pi)^{-1}\int d\omega\,A_k(\omega)$.
Inverting this bound on the stored width-reference spectra at the three
temperatures excludes the locked-reference gap $E_g=0.4929E_{\rm F}$.
At the common probe width $\epsilon=0.00625E_{\rm F}$, all three spectra
exclude a hard gap above $0.076E_{\rm F}$.

For comparison, at the roots $D_{\epsilon=T}$ is $0.950$, $0.944$, and $0.938$ of the
value for a density-matched free gas read with the same kernel. This is
a thermally averaged spectral density near zero energy. In the width
scans, $(Q,\mu)$ is held fixed within each temperature, and
$(D_{0.45}-D_{0.55})/D_{0.50}=0.47$--$0.51\%$ across the three
temperatures. Here $D_\kappa$ denotes $D_{\epsilon=T}$ at the internal
width $\kappa$ in units of $k_{\rm F}$.
Over $\kappa=0.3$--$0.7k_{\rm F}$ at $T=0.025E_{\rm F}$, this midpoint
spectral density decreases monotonically by $1.9\%$
(Sec.~\ref{sec:width-states}).

\subsection{Counting with bare and dressed propagators}\label{sec:countchecks}
\label{sec:closure-count}
\begin{figure}[!t]
\includegraphics[width=\columnwidth]{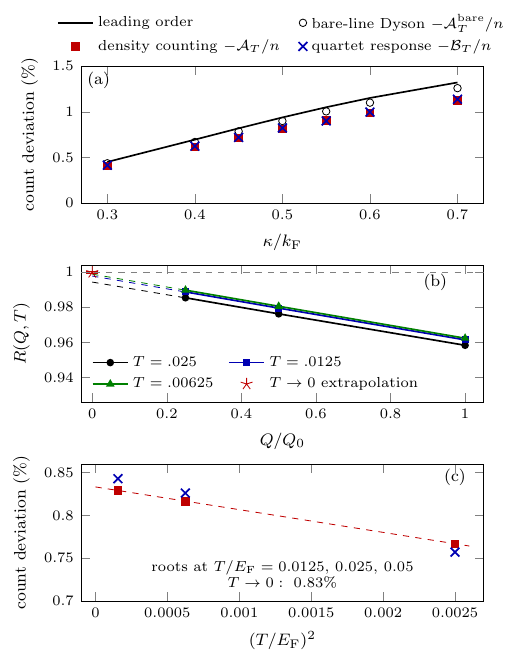}
\caption{One-dimensional consistency checks. (a)~Counting correction versus
$\kappa$ at $T=0.025E_{\rm F}$, $Q=2.638$, and $\mu=1.0055E_{\rm F}$.
The solid line shows the leading expression $-4Q(J_{4p}-J_{4h})/n$.
Open circles show $-\mathcal A_T^{\rm bare}/n$ from the bare-line Dyson
calculation. Squares show density counting $-\mathcal A_T/n$, obtained
from $n(T)$ and $k^*$ after the free-gas thermal subtraction.
Crosses show the quartet response
$-\mathcal B_T/n$ evaluated from the spectra of the same self-consistent
states as the squares.
(b)~$R(Q,T)=\mathcal A_T^{\rm bare}/[4Q(J_{4p}-J_{4h})]$ versus
$Q/Q_0$, comparing the bare-line counting correction with
Eq.~(\ref{eq:J}). Here, $Q_0=2.648$ and $\mu$ is that of the
$T=0.025E_{\rm F}$ root. Dashed segments extrapolate to
$Q=0$. The star then extrapolates to $T\to0$.
Its bar spans the linear and $T^2$ extrapolation laws.
(c)~Squares and crosses compare density counting and the quartet
response for the same roots at density $4k_{\rm F}/\pi$ versus
$(T/E_{\rm F})^2$. The dashed line formally extrapolates the squares
along the unpaired branch to $T=0$. Temperatures are in units of $E_{\rm F}$.}

\label{fig:test}
\end{figure}

We first examine the leading term in Eq.~(\ref{eq:J}) using the
bare-line Dyson approximation. In this approximation the self-energy
is evaluated with free internal lines and the external Dyson equation is
then solved (Appendix~\ref{sec:firstorder}). We compute the density in the
Matsubara representation and obtain $k^*$ from the pole. In
this comparison, $\mu$ is held at the value of the $T=0.025E_{\rm F}$
root while $Q$ and $T$ are varied independently. The free-gas thermal
term $n_{\rm free}(\mu,T)-4k_\mu/\pi$ is removed from the density.
Writing the resulting correction as
$\mathcal A_T^{\rm bare}=n_{\rm bare}(T)-4k^*_{\rm bare}/\pi
-[n_{\rm free}(\mu,T)-4k_\mu/\pi]$, we define
$R(Q,T)=\mathcal A_T^{\rm bare}/[4Q(J_{4p}-J_{4h})]$.
Figure~\ref{fig:test}(b) and Table~\ref{tab:closure} show that $R$ approaches unity. At each temperature,
the ratios at the three amplitudes follow a straight line in $Q$ within
$9\times10^{-5}$. This linear relative deviation is consistent with an
$O(Q^2)$ absolute remainder. Extrapolation to $Q\to0$ gives $0.9944$,
$0.9978$, and $0.9989$ at
$T=0.025$, $0.0125$, and $0.00625E_{\rm F}$, respectively. The remaining
deviation from unity decreases by factors of $2.6$ and $1.9$ per halving of
$T$, which lies between a linear and a $T^2$ law, so that the extrapolation
to $T\to0$ gives $0.9991$ for a $T^2$ law and $1.0006$ for a linear law. At
$T=0.00625E_{\rm F}$, the momentum grid near $k_{\rm F}$ is refined to
$7.5\times10^{-4}k_{\rm F}$, and the same refinement changes the ratio at
$T=0.0125E_{\rm F}$ by only $8\times10^{-5}$.

At density $4/\pi$, the fitted spectral peaks of the roots cross zero
at $k^*=1.00870$, $1.00842$, and $1.00836$. We use these peak crossings as
finite-temperature estimators of the Fermi points and define
\begin{equation}
\mathcal A_T=n(T)-\frac{4k^*(T)}{\pi}
-\left[n_{\rm free}(\mu,T)-\frac{4\sqrt\mu}{\pi}\right],
\label{eq:AT}
\end{equation}
which gives $-0.00977$, $-0.01040$, and $-0.01056$ at $T=0.05$, $0.025$,
and $0.0125E_{\rm F}$. A $T^2$ fit gives $\mathcal A_0=-0.0106$, or
$-0.83\%$ of the density. Thus the extrapolated Fermi-volume count is
$0.83\%$ larger than the density on the formal unpaired continuation.
The pairing response discussed in Sec.~\ref{sec:pairing} can preempt this
continuation.

We evaluate the zero-temperature spectral expression (\ref{eq:dressed})
on the stored finite-temperature spectra
$A_k(\omega)=\sum_jw_j\delta(\omega-e_j)$, and denote the resulting
quartet response by $\mathcal B_T$. These spectra belong to the same self-consistent states
used for density counting $\mathcal A_T$ in Eq.~(\ref{eq:AT}). The
squares and crosses in Fig.~\ref{fig:test}(a) and (c) therefore
compare two calculations on each state. At a stationary zero-temperature
solution, both quantities reduce to $4n_Q$. Their finite-temperature
comparison retains thermal corrections after the free-gas subtraction
in Eq.~(\ref{eq:AT}).

For a fixed spectrum $A_T$, the density with finite-temperature
occupations differs from that with zero-temperature occupations by
\begin{equation}
\begin{aligned}
n_T[A_T]-n_0[A_T]&=4\int\frac{dk}{2\pi}\int d\omega\,A_T(k,\omega)\\
&\quad\times[f_T(\omega)-\Theta(-\omega)].
\end{aligned}
\end{equation}
Subtracting the free-gas counterpart leaves the integral with
$A_T-A_{\rm free}$. This interacting thermal contribution, together
with the temperature dependence of the self-energy and the peak
crossing, enters the comparison of $\mathcal A_T$ and $\mathcal B_T$.

Table~\ref{tab:direct} gives the results on the full momentum grid.
The ratio $\mathcal B_T/\mathcal A_T$ ranges from $0.987$ to $1.022$ for the three roots and seven width
states on $N_k=4097$. For the $T=0.025E_{\rm F}$ root and width states,
the bare-line Dyson counting correction has a $7$--$12\%$ larger magnitude
than its self-consistent counterpart.
A formal $T^2$ extrapolation of the two lower-temperature spectral
responses gives $\mathcal B_0=-0.0108$, close to the density-based
estimate $\mathcal A_0=-0.0106$. The momentum-grid and quadrature
comparisons are given in Appendix~\ref{app:count}.
The observed agreement supports the sign and width dependence of the
counting anomaly with dressed lines.

\begin{table*}[t]
\caption{Density counting and quartet response for the same
self-consistent states. The zero-temperature expression
$\mathcal B_T$ is evaluated on their stored finite-temperature spectra,
while $\mathcal A_T$ is obtained from $n(T)$ and $k^*$ through
Eq.~(\ref{eq:AT}). The ratio is $\mathcal B_T/\mathcal A_T$.
The last column gives the bare-line Dyson $\mathcal A_T$ divided by
its self-consistent value at the same $(Q,\mu,T)$.}
\label{tab:direct}
\begin{ruledtabular}
\begin{tabular}{llcccc}
state & $\kappa/k_{\rm F}$ & $\mathcal B_T$ & $\mathcal A_T$ & ratio & bare / s.c.\\
\hline
root, $T=0.05$ & $0.50$ & $-0.00964$ & $-0.00977$ & $0.987$ & ---\\
root, $T=0.025$ & $0.50$ & $-0.01052$ & $-0.01040$ & $1.012$ & $1.107$\\
root, $T=0.0125$ & $0.50$ & $-0.01073$ & $-0.01056$ & $1.016$ & ---\\
$T=0.025$, $Q=2.638$ & $0.30$ & $-0.00530$ & $-0.00518$ & $1.022$ & $1.073$\\
$T=0.025$, $Q=2.638$ & $0.40$ & $-0.00796$ & $-0.00782$ & $1.017$ & $1.091$\\
$T=0.025$, $Q=2.638$ & $0.45$ & $-0.00918$ & $-0.00913$ & $1.005$ & $1.098$\\
$T=0.025$, $Q=2.638$ & $0.50$ & $-0.01050$ & $-0.01038$ & $1.011$ & $1.104$\\
$T=0.025$, $Q=2.638$ & $0.55$ & $-0.01146$ & $-0.01154$ & $0.994$ & $1.109$\\
$T=0.025$, $Q=2.638$ & $0.60$ & $-0.01271$ & $-0.01259$ & $1.010$ & $1.114$\\
$T=0.025$, $Q=2.638$ & $0.70$ & $-0.01450$ & $-0.01433$ & $1.012$ & $1.120$\\
\end{tabular}
\end{ruledtabular}
\end{table*}

\subsection{Width dependence and low-energy thermodynamics}
\label{sec:width-states}

We next examine how the quartet width affects the count and the
low-energy spectrum. In Fig.~\ref{fig:test}(a), we hold $Q$, $\mu$, and
$T$ fixed while varying $\kappa$. The form factor retains its peak
normalization $\phi(0,0)=1$. We denote by $s_Q$ and $s_N$ the fermionic
entropy densities of the quartet branch and the free gas at the same
reference density and temperature, respectively. The entropy deficit in
the width scan is estimated from the quasiparticle velocity, as described
below.
Tables~\ref{tab:width} and \ref{tab:dressing} compare the widths at
fixed $(Q,\mu)$, where $Q=2.638$ lies $0.4\%$ below the root amplitude.
Each width is solved self-consistently at these common parameters. For
$\kappa=0.3$--$0.7k_{\rm F}$, the self-consistent entropy deficit stays
within $2.25$--$2.60\%$, whereas the Fermi-point shift $k^*-k_\mu$ and
the magnitude of $\mathcal A_T$ change by factors of $4.7$ and $2.8$.
The self-consistent entropy deficit differs from the bare-line result
by $-0.05$ to $+0.11$ percentage
points, comparable to the estimator spread of $0.01$--$0.11$ points.
The tunneling density of states, defined below, is $0.5$--$0.8\%$ above the bare value,
and its midpoint counterpart decreases monotonically by $1.9\%$ across
the widths. The ratio of the self-consistent to bare-line counting
correction follows each width's $Z^3$ within $1\%$.
The shift and pair eigenvalue ratios follow $Z^2$ within about $0.7\%$
and $0.3\%$, respectively. These empirical dressing relations apply to
the sampled $Z=0.961$--$0.978$ interval.

\begin{figure}[!t]
\includegraphics[width=\columnwidth]{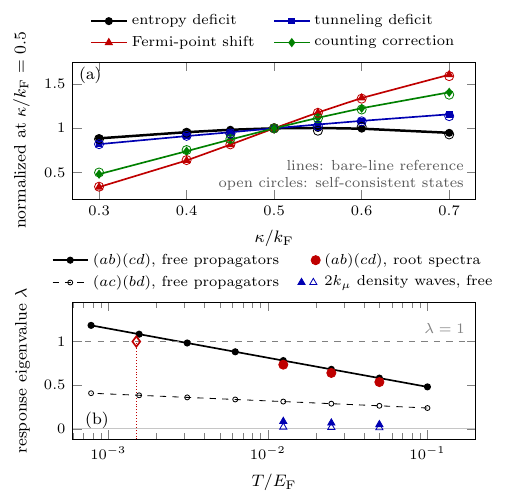}
\caption{Width-dependent observables and linear pairing tests.
(a)~Entropy deficit $1-s_Q/s_N$, tunneling deficit $1-N_{\rm tun}/N_0$
($N_0$: free density of states), Fermi-point shift $k^*-k_\mu$, and
counting correction at the $(Q,\mu,T)$ of Fig.~\ref{fig:test}(a).
Each series is normalized to its own value at $\kappa=0.5k_{\rm F}$.
Lines with filled symbols use the bare-line Dyson approximation for the
first three quantities and $-4Q(J_{4p}-J_{4h})$
[Eq.~(\ref{eq:J})] for the count. Open circles use self-consistent
states, with density counting $-\mathcal A_T$ for the count.
(b)~Black solid and dashed curves show the $(ab)(cd)$ and $(ac)(bd)$ pair
channels with free propagators at the fixed $(Q,\mu)$ of the
$T=0.025E_{\rm F}$ root. Red circles show $(ab)(cd)$ pairing from the
stored root spectra. Blue filled and open triangles show $ab$ and $ac$
density-wave estimates at $P=2k_\mu$, using free propagators at each
root's parameters, where $k_\mu=\sqrt\mu$ is the free-band Fermi momentum. The gray dashed line marks $\lambda=1$. The red open
diamond at $\lambda=1$ and its vertical dotted guide mark
$T_{\rm pair}\approx1.5\times10^{-3}E_{\rm F}$, estimated by extrapolating
the $Z^2$ dressing inferred from the stored root spectra for the
$(ab)(cd)$ pair channel. This is an extrapolated linear instability
scale, rather than a computed phase boundary.}
\label{fig:window}
\end{figure}

Figure~\ref{fig:window}(a) shows that the Fermi-point shift and the
counting correction respond more strongly to a change in width than the
entropy does. The correction in Eq.~(\ref{eq:dressed}) samples the
four-fermion continuum. The leading entropy, on the other hand, depends
on the slope of the quasiparticle dispersion close to the Fermi points.

\begin{table*}[t]
\caption{Width dependence at fixed $(Q,\mu)$. Here, $T=0.025E_{\rm F}$,
$Q=2.638$, $\mu=1.00549E_{\rm F}$, and $k_\mu=1.00274k_{\rm F}$, and the
deficit is defined as $1-2/v^*$. The zero-energy tunneling density of
states $N_{\rm tun}$ is measured relative to its free-gas reference $N_0$.
In each entry with two values, the first
is obtained with the bare-line Dyson reference and the second with
self-consistent lines. The
last three columns give the zero-temperature leading expression
(\ref{eq:J}), and $\mathcal A_T$ from the bare-line Dyson approximation and the
self-consistent states, respectively.}
\label{tab:width}
\begin{ruledtabular}
\begin{tabular}{lcccccc}
$\kappa/k_{\rm F}$ & deficit (\%) & $N_{\rm tun}/N_0$ & $k^*/k_{\rm F}$ & Eq.~(\ref{eq:J}) & bare Dyson & s.c.\\
\hline
$0.30$ & $2.218$ / $2.250$ & $0.9510$ / $0.9556$ & $1.00476$ / $1.00466$ & $-0.00575$ & $-0.00556$ & $-0.00518$\\
$0.40$ & $2.393$ / $2.433$ & $0.9455$ / $0.9505$ & $1.00659$ / $1.00638$ & $-0.00886$ & $-0.00853$ & $-0.00782$\\
$0.45$ & $2.461$ / $2.407$ & $0.9429$ / $0.9490$ & $1.00767$ / $1.00738$ & $-0.01044$ & $-0.01003$ & $-0.00913$\\
$0.50$ & $2.503$ / $2.561$ & $0.9403$ / $0.9459$ & $1.00878$ / $1.00842$ & $-0.01196$ & $-0.01146$ & $-0.01038$\\
$0.55$ & $2.513$ / $2.491$ & $0.9377$ / $0.9445$ & $1.00986$ / $1.00941$ & $-0.01338$ & $-0.01280$ & $-0.01154$\\
$0.60$ & $2.492$ / $2.604$ & $0.9352$ / $0.9416$ & $1.01086$ / $1.01032$ & $-0.01468$ & $-0.01402$ & $-0.01259$\\
$0.70$ & $2.375$ / $2.383$ & $0.9308$ / $0.9384$ & $1.01245$ / $1.01177$ & $-0.01684$ & $-0.01604$ & $-0.01433$\\
\end{tabular}
\end{ruledtabular}
\end{table*}

\begin{figure}[t]
\includegraphics[width=\columnwidth]{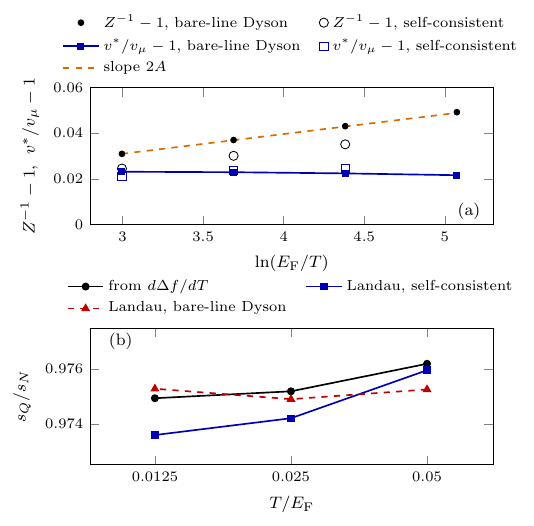}
\caption{Light cone and entropy in one dimension. (a)~Black circles show
$Z^{-1}-1$, and blue squares show $v^*/v_\mu-1$, where $v_\mu=2\sqrt\mu$
is evaluated at the chemical potential of each plotted state. Filled
symbols use the bare-line Dyson approximation at the fixed $(Q,\mu)$ of
the $T=0.025E_{\rm F}$ root. The orange dashed line has the predicted
slope $2A$ and is anchored to the highest-temperature filled circle.
Open black circles and open blue squares show the corresponding quantities
for the self-consistent roots, whose $Q$ grows by $7\%$ as $T$ decreases.
The bare-line velocity ratio changes by less than $0.1\%$ per halving of
$T$. (b)~Entropy ratio obtained from $d\Delta f/dT$, from the Landau term
of the self-consistent quasiparticle, and from that of the bare-line
Dyson approximation.}
\label{fig:lightcone}
\end{figure}

\paragraph{Fermionic entropy and tunneling density of states.} Since $\partial\Omega/\partial Q=0$ at a stationary root, one has
$d\Delta f/dT=-(s_Q-s_N)$, where $s_N=(2\pi/3)T$ is the entropy of the free
gas at the same density up to a $T^3$ term. A fit
$\Delta f=\Delta f_0+\beta_2T^2+\beta_4T^4$ through the three roots on $N_k=4097$ gives
the local values $s_Q/s_N=0.9762$, $0.9752$, and $0.9749$, while the step
$T=0.05\to0.025$ gives $0.9756$ on both grids. On the other hand, the
leading Landau quasiparticle contribution to the entropy~\cite{CarneiroPethick1975}
reads
\begin{equation}
\begin{aligned}
s&=\frac{\pi^2}{3}\sum_iN_i^*(0)\,T,\qquad N^*(0)=\frac{1}{\pi v^*},\\
v^*&=Z\,(v_k+\partial_k{\rm Re}\,\Sigma)_{k^*,0},
\end{aligned}
\label{eq:landau}
\end{equation}
where $N_i^*(0)$ is the quasiparticle density of states of component $i$.
The tunneling density of states per component is
$N_{\rm tun}=\int dk\,A(k,0)/(2\pi)$, with $\omega=0$ at the chemical
potential. The residue $Z$ measures the single-particle weight of a
quasiparticle peak. For a low-energy pole $A(k,\omega)\simeq Z\delta(\omega-E_k^*)$,
its contribution is $N_{\rm tun}\simeq Z/(\pi v^*)$, whereas
$N^*(0)=1/(\pi v^*)$ counts the quasiparticle levels entering the entropy.
The Landau entropy expression uses only the stored spectra. The quasiparticle peak at each
momentum is located from the broadened spectral function or from the
first moment of the spectral weights near the dominant pole, and $v^*$ is the slope of
a linear fit over $|E^*|<0.02$--$0.04E_{\rm F}$. As summarized in
Table~\ref{tab:entropy}, the two routes agree within $1.4\times10^{-3}$ of
the entropy. The entropy deficit inferred from the spectrum is larger by about
$10^{-3}$ at the lower temperatures. The two routes are compared in Fig.~\ref{fig:lightcone}(b). We consider
only the fermionic entropy of the static quartet description.

\begin{table*}[t]
\caption{Entropy and quasiparticle parameters of the roots on $N_k=4097$. Here,
$N_{\rm tun}/N_0=v_{\rm F}/(v_{k^*}+\partial_k{\rm Re}\,\Sigma)$ is the
tunneling density of states relative to the free value $N_0$.
Both $N_0$ and $v_{\rm F}=k_{\rm F}/m=2$ refer to the free gas at the fixed
reference density.}
\label{tab:entropy}
\begin{ruledtabular}
\begin{tabular}{lcccccc}
$T/E_{\rm F}$ & $k^*/k_{\rm F}$ & $v^*/v_{\rm F}$ & $2/v^*$ & $s_Q/s_N$ from $\Delta f$ & $Z$ & $N_{\rm tun}/N_0$\\
\hline
$0.05$ & $1.0087$ & $1.0246$ & $0.9760$ & $0.9762$ & $0.976$ & $0.953$\\
$0.025$ & $1.0084$ & $1.0265$ & $0.9742$ & $0.9752$ & $0.971$ & $0.946$\\
$0.0125$ & $1.0084$ & $1.0271$ & $0.9736$ & $0.9749$ & $0.966$ & $0.941$\\
\end{tabular}
\end{ruledtabular}
\end{table*}

\paragraph{Low-energy self-energy.}
\label{sec:lightcone}
In the following derivation, $k_{\rm F}$ and $v_{\rm F}$ denote the Fermi momentum
and velocity of the internal lines, namely $k_\mu=\sqrt\mu$ and
$v_\mu=2\sqrt\mu$ for free lines at $\mu$. Near $k=k_{\rm F}+\delta k$, the
dominant sector of $S(k,\nu)=-{\rm Im}\,\Sigma^R(k,\nu)/\pi$ has $u,v\to0$,
where line 3 is near $+k_{\rm F}$ and lines 2 and 4 are near
$-k_{\rm F}$. When the internal dispersions are linearized with the
velocity $v_{\rm F}$, the delta function of $S$ at the frequency $\nu$
fixes $u$, and one obtains
\begin{equation}
\begin{aligned}
\xi_3&=-\tfrac12(\nu+v_{\rm F}\delta k),\\
\xi_2+\xi_4&=-\tfrac12(\nu-v_{\rm F}\delta k),\\
\xi_2-\xi_4&=-\sqrt2\,v_{\rm F}v ,
\end{aligned}
\end{equation}
where $v$ is the Jacobi momentum of Eq.~(\ref{eq:vertex}).
At $T=0$, only three holes or three particles contribute, so that the
free difference ranges over an interval of length $|\nu-v_{\rm F}\delta k|$,
which is open only when $\nu+v_{\rm F}\delta k$ and $\nu-v_{\rm F}\delta k$
have the same sign. Including the two sectors in which the line 2 or 4
sits near $+k_{\rm F}$, each weighted by $e^{-k_{\rm F}^2/\kappa^2}$, one has
\begin{equation}
\begin{aligned}
S(k_{\rm F}+\delta k,\nu)&=A\,|\nu-v_{\rm F}\delta k|\,
\Theta(|\nu|-v_{\rm F}|\delta k|),\\
A&=\frac{Q(1+2e^{-k_{\rm F}^2/\kappa^2})}{16\pi^2v_{\rm F}^2}.
\end{aligned}
\label{eq:lc}
\end{equation}
Between $T$ and an upper scale $\Lambda$, the Kramers-Kronig integrals at
the Fermi point give
$\partial_\omega{\rm Re}\,\Sigma=-2A\ln(\Lambda/T)+b_\omega$ and
$\partial_k{\rm Re}\,\Sigma=+2Av_{\rm F}\ln(\Lambda/T)+b_k$, where $b_\omega$
and $b_k$ are the regular remainders. The step function removes
an interval of width $2v_{\rm F}|\delta k|$, which lies inside the thermal
window and adds no large logarithm. The external bare-band velocity is
$v_0=k^*/m$, distinct from the velocity of the internal lines. With free
internal lines, the linearized pole equation gives
$v^*=[v_0+b_k+2Av_{\rm F}\ln(\Lambda/T)]/
[1-b_\omega+2A\ln(\Lambda/T)]$.
For self-consistent pole-like internal lines, replace $(k_{\rm F},v_{\rm F})$
by $(k^*,v^*)$ in the light cone. Its coefficient becomes
$A_{\rm sc}=QZ^3[1+2e^{-(k^*)^2/\kappa^2}]/(16\pi^2(v^*)^2)$, so that
\begin{equation}
\begin{aligned}
v^*\bigl[1-b_\omega+2A_{\rm sc}\ln(\Lambda/T)\bigr]
&=v_0+b_k+2A_{\rm sc}v^*\ln(\Lambda/T),\\
v^*&=\frac{v_0+b_k}{1-b_\omega}.
\end{aligned}
\end{equation}
Within this linearized self-consistent treatment, the leading logarithm
enters $Z$ and $N_{\rm tun}$ but cancels in $v^*$. In
the bare-line Dyson approximation at the fixed $(Q,\mu)$ of the $T=0.025E_{\rm F}$ root,
halving $T$ from $0.05$ to $0.00625E_{\rm F}$ raises $Z^{-1}$ by $0.00606$,
$0.00607$, and $0.00613$, in comparison with $2A\ln2=0.00599$. At the
same time, it lowers $N_{\rm tun}$ by $0.57$, $0.53$, and $0.51\%$ and
changes $v^*$ by only $0.02$, $0.05$, and $0.08\%$. For the self-consistent
roots, one has $Z^{-1}-1=0.025$, $0.030$, and $0.035$, which rise by
$0.0056$ and $0.0050$ per halving, in comparison with $2A\ln2=0.0051$, where
$A=A_{\rm sc}$. The
amplitude of these roots grows by $7\%$ as $T$ decreases, so that their
$v^*$ does not test the cancellation at fixed $(Q,\mu)$. These results are
shown in Fig.~\ref{fig:lightcone}(a).

\subsection{Pairing and density-wave responses}
\label{sec:pairing}
\label{sec:pairing-response}
With the pair amplitudes $F_{xy}(p)=\langle y_{-p}x_p\rangle$, the Wick part
of Eq.~(\ref{eq:HD}) reads
$\Delta^*\phi\,(F_{ab}F_{cd}-F_{ac}F_{bd}+F_{ad}F_{bc})$, as in the treatment of competing pairing in Ref.~\cite{Jiang2017}. This
is a single-vertex term of the Nambu-extended functional and vanishes on the
quartet-restricted branch. It acts as a pairing field on each pair
channel,
$D_{ab}(p)=\Delta^*\int\frac{d^dp'}{(2\pi)^d}\phi_{ab}(p,p')F_{cd}(p')$, and
the linear response $|F_{ab}|=\chi|D_{ab}|$ closes the system
\begin{equation}
\begin{aligned}
F_{ab}&=-\chi\,\Delta\,\mathcal K F^*_{cd},\qquad
F_{cd}=-\chi\,\Delta\,\mathcal K F^*_{ab},\\
\chi(p)&=T\sum_nG_p(i\omega_n)G_{-p}(-i\omega_n),
\end{aligned}
\end{equation}
where $\mathcal K=\phi_{ab}/(2\pi)^d$. Within this first-order pairing kernel, the branch has a linear
instability when the largest eigenvalue $\lambda$ of $\sqrt Q\,\chi^{1/2}\mathcal K\chi^{1/2}$
reaches unity. The
kernels are given by
$\phi_{ab}(p,p')=\phi_{ad}(p,-p')=e^{-|p-p'|^2/(8\kappa^2)}$ and
$\phi_{ac}=e^{-(p^2+p'^2)/(4\kappa^2)}$. Since $\phi_{ab}(p,p')=\phi_{ad}(p,-p')$,
the $(ab)(cd)$ and $(ad)(bc)$ channels are degenerate, so that the linear problem has a degenerate leading subspace.
The linear response determines the onset of the instability. Nonlinear
terms are needed to determine the resulting pair order. For $F=\chi h$ with $h$ smooth on
the scale $\kappa$, we discretize the problem with the cell integrals of
$\chi$ resolved below $T/v_{\rm F}$. On the solver grid, $\chi$ is computed
from the stored spectral weights as
$\sum_{jl}w_jw_l(1-f_j-f_l)/(e_j+e_l)$, with $f_j=f(e_j)$. The results
are summarized in Table~\ref{tab:pair}.

\begin{table}[t]
\caption{Largest pair eigenvalues. The dressed values use the stored
self-consistent spectra, and their ratio to the free-line values is close
to $Z^2$.}
\label{tab:pair}
\begin{ruledtabular}
\begin{tabular}{lccccc}
$T/E_{\rm F}$ & $\lambda_{ab}$ free & $\lambda_{ab}$ dressed & ratio & $\lambda_{ac}$ free & $Z^2$\\
\hline
$0.05$ & $0.565$ & $0.535$ & $0.948$ & $0.256$ & $0.953$\\
$0.025$ & $0.681$ & $0.640$ & $0.939$ & $0.288$ & $0.942$\\
$0.0125$ & $0.787$ & $0.735$ & $0.934$ & $0.314$ & $0.933$\\
\end{tabular}
\end{ruledtabular}
\end{table}

At the fixed $(Q,\mu)$ of the $T=0.025$ root, the free-propagator
$\lambda_{ab}$ grows by $0.101$ per halving of $T$, and $\lambda_{ac}$ grows
by $0.024$. The free-propagator crossing $\lambda_{ab}=1$ occurs at
$2.8\times10^{-3}E_{\rm F}$. To estimate the dressed crossing, we use the ratios
$r(T)=\lambda_{ab}^{\rm dressed}/\lambda_{ab}^{\rm free}$ at matched
$(Q,\mu,T)$ in Table~\ref{tab:pair}. They decrease from $0.948$ to
$0.939$ and $0.934$, about $0.007$ per halving of temperature, close to
the observed $Z^2$ trend. We continue this reduction linearly in
$\ln(1/T)$ and multiply it by the free-line response at fixed $(Q,\mu)$.
Solving for the crossing of this estimated response with unity gives
$T_{\rm pair}\approx1.5\times10^{-3}E_{\rm F}$.
The estimate extrapolates the spectral reduction by a factor of eight
below the lowest computed temperature. The red diamond in Fig.~\ref{fig:window}(b) marks this
extrapolated scale at $\lambda=1$. At the
fixed $(Q,\mu)$ of Table~\ref{tab:width}, the dressed eigenvalues of the
self-consistent states at $T=0.025E_{\rm F}$ are $\lambda_{ab}=0.49$--$0.77$
for $\kappa=0.3$--$0.7k_{\rm F}$, which are the free-line values multiplied
by $Z^2$ within $0.3\%$ (Table~\ref{tab:dressing}). For free lines,
$\lambda$ is proportional to $\sqrt Q$. On a
rotation-invariant Fermi surface, the positive kernel makes the leading
channel isotropic, and the logarithm has the prefactor
$\sqrt Q\,N(0)\langle\phi\rangle_{\rm FS}$, where $N(0)$ is the
density of states of one component. To leading logarithmic accuracy,
\begin{equation}
T_{\rm pair}=\Lambda_\kappa\exp\Bigl[-\frac{1}{\sqrt Q\,N(0)\,
\langle\phi\rangle_{\rm FS}}\Bigr],
\label{eq:Tpair}
\end{equation}
where $\Lambda_\kappa$ is a vertex-dependent energy cutoff. Evaluating
the angular average of $\phi_{ab}$ on the free Fermi surface gives
\begin{equation}
\langle\phi\rangle_{\rm FS}=
\begin{cases}
(1+e^{-2\eta})/2, & d=1,\\
e^{-\eta}I_0(\eta), & d=2,\\
(1-e^{-2\eta})/(2\eta), & d=3,
\end{cases}
\end{equation}
Here $\eta=k_\mu^2/(4\kappa^2)$ for the free reference band and $I_0$ is
the modified Bessel function of the first kind. For $d\ge2$, the leading
channel favors nearly collinear conversion at small width, with
$\langle\phi\rangle_{\rm FS}\propto(\kappa/k_\mu)^{d-1}$.
As shown in Table~\ref{tab:pairdim}, the radial $s$-wave problem with the
angle-averaged kernel confirms this within $1.7\%$. At
$T=0.025E_{\rm F}$ in one dimension, we obtain $\lambda_{ab}=0.51$, $0.59$,
$0.68$, $0.76$, $0.84$, $0.99$, and $1.12$ for $\kappa=0.3$, $0.4$, $0.5$,
$0.6$, $0.7$, $1.0$, and $1.5k_{\rm F}$, respectively. For
$\kappa=0.3$--$1.0k_{\rm F}$, the free lines give
$T_{\rm pair}=5.4\times10^{-4}$, $1.2\times10^{-3}$, $2.8\times10^{-3}$,
$5.8\times10^{-3}$, $9.8\times10^{-3}$, and $0.024E_{\rm F}$.

\begin{figure}[t]
\includegraphics[width=\columnwidth]{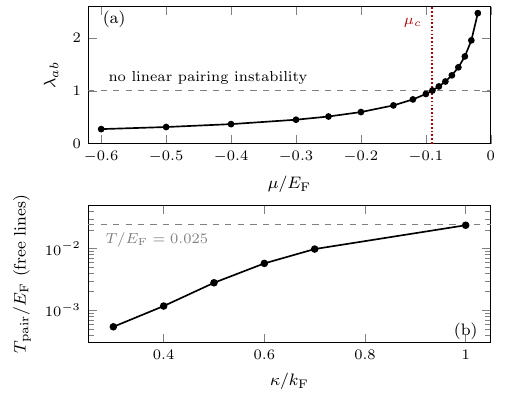}
\caption{Free-propagator pairing thresholds. (a)~Pair eigenvalue at
$T=10^{-4}E_{\rm F}$ as a function of negative $\mu$, where the free band
has no Fermi sea,
obtained with free lines at $Q_0$ and $\kappa=0.5k_{\rm F}$. This criterion predicts no linear pairing instability below $\mu_c$. (b)~$T_{\rm pair}$ as a function of $\kappa$,
obtained with free lines at the $(Q,\mu)$ of the $T=0.025E_{\rm F}$ root.}
\label{fig:pairS}
\end{figure}

For $\mu<0$, the free reference band has no Fermi sea and
$\xi_p=p^2/(2m)+|\mu|>0$. Its pair bubble
$\chi_0(p,T)=\tanh(\xi_p/2T)/(2\xi_p)$ tends to $1/(2\xi_p)$ as $T\to0$,
so the momentum integral has no Cooper logarithm.
At $Q_0$ and $\kappa=0.5k_{\rm F}$, one obtains
$\lambda_{ab}(T\to0)=0.27$, $0.45$, $0.60$, $0.94$, $1.45$, and $2.48$
for $\mu=-0.6$, $-0.3$, $-0.2$, $-0.1$, $-0.05$, and $-0.02E_{\rm F}$,
respectively. Finite eigenvalues can still exceed unity.
In this free-line calculation, the zero-temperature instability occurs
for $\mu>\mu_c$, where
$\mu_c=-0.072$, $-0.091$, and
$-0.104E_{\rm F}$ for $\kappa=0.3$, $0.5$, and $0.7k_{\rm F}$, respectively.
At $\kappa=0.5k_{\rm F}$, one has $\mu_c=-0.029$ and $-0.26E_{\rm F}$ at $Q_0/4$
and $4Q_0$, respectively. These values are obtained with free lines without
the $O(Q)$ band shift, and they are shown in Fig.~\ref{fig:pairS}(a)
together with the width dependence of $T_{\rm pair}$.

\paragraph{Density waves.}
A density modulation in component $a$ has
$\delta n_a(x)=\eta_a\cos(Px+\theta_a)$ and a particle--hole coherence
$\langle a^\dagger_{k+P}a_k\rangle$.
We evaluate the response at $P=2k_\mu$, with $k_\mu=\sqrt\mu$, which
connects the two Fermi points of the free reference band used in this
calculation.
Since $\Sigma_a$ does not depend on $G_a$, the four-line term has no
intra-component kernel. A density wave of the wavevector $P$ in the
component $c$ induces that of the wavevector $-P$ in the component $a$
through
\begin{equation}
\mathcal K_{ac}(k_1,k_3)=-Q\int\frac{dk_2}{2\pi}\,\phi(k)\,\phi(k')\,
\frac{1-f_2-f_4}{\xi_2+\xi_4},
\end{equation}
where $k'$ denotes the momenta after the transfer, and similarly for the
pairs $(ab)$ and $(ad)$. Using the static kernel and free lines at the $(Q,\mu,T)$ of the roots, we
obtain the largest eigenvalues of the linearized problem with the static
Lindhard factors as $\lambda_{ac}=0.012$, $0.016$, and
$0.020$ and $\lambda_{ab}=0.048$, $0.066$, and $0.084$ at the three roots,
which are converged in the grid. At $\kappa=k_{\rm F}$, they are $0.16$ and
$0.25$ at $T=0.025E_{\rm F}$.

The eigenvalues at the computed roots remain below unity in both
responses. The growth of the pair eigenvalue on cooling indicates that
pair order can preempt the formal unpaired continuation of
Sec.~\ref{sec:countchecks}. The counting relation in
Eq.~(\ref{eq:general}) is derived for the branch with $F_{xy}=0$.
To follow a paired branch, one must include anomalous propagators.

\section{Summary and perspectives}\label{sec:summary}

In this paper, we have investigated the relation between the density
and the Fermi-volume count in a four-component fermion system with quartet condensation.
The frequency shift at a quartet vertex gives a response of the
Luttinger-Ward functional to the phase winding of the quartet field.
We have evaluated this response at leading order as a weighted
difference between four-particle and four-hole contributions. The
thermodynamic derivation identifies the same difference as the part of
the density change due to continuum occupation after the motion of the
Fermi boundary is accounted for. In the Gaussian model, band curvature
and radial phase space favor opposite signs of the correction in one
and three dimensions. In two dimensions the sign depends on the cutoff.

We have also compared the density count with the spectral response
on the same low-temperature self-consistent states. The two
calculations support the sign and width dependence of the counting
correction. At fixed $Q$ and peak vertex normalization, its magnitude
changes by a factor of $2.8$ over the studied width interval, whereas
the fermionic entropy deficit remains close to $2.5\%$. The
difference follows from the quantities sampled in the two calculations.
The count is sensitive to the four-fermion continuum, and the
leading entropy depends on the quasiparticle velocity near the Fermi
points.

The one-dimensional calculation assumes a homogeneous phase-coherent
quartet field. In strict one dimension, thermal phase fluctuations
limit its coherence~\cite{Hohenberg1967}. The stiffness estimate in
Appendix~\ref{sec:phase} gives a classical coherence length of about
$2$--$4k_{\rm F}^{-1}$ at the two lower temperatures. Intertube coupling
could stabilize phase coherence, providing a setting for the
homogeneous calculation. It would be interesting to examine how phase
fluctuations and the onset of pair order change the density-volume
relation. This is particularly relevant to attractive
one-dimensional four-component systems, where quartet phases with
gapped fermions have been found~\cite{Lecheminant2005,Capponi2008}.

In a phase-coherent four-component Fermi gas, momentum-resolved
spectroscopy~\cite{CazalillaRey2014,Stewart2008} could be used to locate
the Fermi points, while thermodynamic measurements probe the entropy.
Comparing these observables would help distinguish changes in
continuum occupation from the low-energy dispersion of the fermions.

\begin{acknowledgments}
Y.G. is supported by the RIKEN Special Postdoctoral Researchers Program.
\end{acknowledgments}

\appendix

\section{Bare-line reference calculations}
\label{sec:firstorder}
The self-energy uses free internal lines and is linear in $Q$, but Dyson
resummation of the external propagator retains higher powers. This
Dyson approximation is therefore distinct from a strict first-order
expansion. Its counting limits are tested in Sec.~\ref{sec:closure-count}.

\subsection{Bare-line Dyson approximation}
Replacing the internal lines of Eq.~(\ref{eq:sigma}) by free lines at the
$(Q,\mu,T)$ of a root defines the bare-line Dyson approximation
$G=1/(G_0^{-1}-\Sigma^{(1)}[G_0])$. Thus the external propagator contains
interaction effects, while the three internal lines remain free.
The spectral function of its self-energy,
$S=-{\rm Im}\,\Sigma^{R}/\pi$, follows exactly from the delta function. With
$E_3=\xi_{k_2}+\xi_{k_3}+\xi_{k_4}=3(u-u_c)^2+A_{\min}+v^2$, where
$u_c=2\sqrt2k/3$ and $A_{\min}=k^2/3-3\mu$ for $m=1/2$, the substitution
$u=u_c+D\sin\vartheta$ and $v=\pm\sqrt3D\cos\vartheta$ with
$D^2=(-\nu-A_{\min})/3$ gives
\begin{equation}
S(k,\nu)=\frac{Q}{8\sqrt3\pi^2}\int_{-\pi/2}^{\pi/2}d\vartheta\sum_\pm
e^{-(u^2+v^2)/(4\kappa^2)}\,\mathcal N_{bcd}\,\Theta_K ,
\label{eq:S}
\end{equation}
where $\Theta_K=\prod_{j=2}^4\Theta(K-|k_j|)$ enforces the internal-leg
cutoff, with the external momentum restricted to $|k|\le K$.
For $\nu>-A_{\min}$ there is no real energy shell and $S(k,\nu)=0$.
The real part is obtained by the
Hilbert transform, and the pole $E^*_k=\xi_k+{\rm Re}\,\Sigma^{(1)}(k,E^*_k)$
gives $k^*$, the quasiparticle velocity $v^*$, and the residue $Z$.
We denote the on-shell self-energy damping scale by
$\Gamma=-{\rm Im}\,\Sigma^R(k^*,0)$.
Direct two-dimensional momentum integration of Eq.~(\ref{eq:sigma})
at Matsubara frequencies reproduces the spectral route.

\subsection{Comparison with the roots}
Table~\ref{tab:firstorder} compares the bare-line Dyson approximation with the
self-consistent roots. The bare-line quasiparticle entropy ratio agrees
with the thermodynamic ratio inferred from $\Delta f$ to about $10^{-3}$.
The Fermi point $k^*$ differs by less than $4\times10^{-4}k_{\rm F}$. At $k^*$ and
$T=0.025E_{\rm F}$, the self-consistent Matsubara self-energy is the bare
one multiplied by $0.934$--$0.936$ for the real part and by $0.900$--$0.911$
for the imaginary part at $i\pi T$ and $3i\pi T$, while $Z^3=0.915$.

\begin{table*}[t]
\caption{Bare-line Dyson approximation in comparison with the self-consistent
roots. The entropy densities $s_Q$ and $s_N$ refer to the quartet branch
and the free gas at the same density and temperature.}
\label{tab:firstorder}
\begin{ruledtabular}
\begin{tabular}{lccccccc}
& \multicolumn{4}{c}{bare-line Dyson} & \multicolumn{2}{c}{self-consistent} & from $\Delta f$\\
$T/E_{\rm F}$ & $k^*/k_{\rm F}$ & $2/v^*$ & $Z$ & $\Gamma/T$ & $k^*/k_{\rm F}$ & $2/v^*$ & $s_Q/s_N$\\
\hline
$0.05$ & $1.00908$ & $0.97525$ & $0.9715$ & $0.0255$ & $1.00870$ & $0.97595$ & $0.97619$\\
$0.025$ & $1.00881$ & $0.97490$ & $0.9643$ & $0.0268$ & $1.00842$ & $0.97420$ & $0.97518$\\
$0.0125$ & $1.00872$ & $0.97528$ & $0.9581$ & $0.0264$ & $1.00836$ & $0.97359$ & $0.97493$\\
\end{tabular}
\end{ruledtabular}
\end{table*}

\subsection{Amplitude and cutoff}
At fixed $\mu$, $T=0.025E_{\rm F}$, and $\kappa=0.5k_{\rm F}$, the deficit
$1-2/v^*$ after subtracting the contribution $1-1/\sqrt\mu=0.27\%$ from $\mu$ alone is $0.00900$,
$0.00882$, $0.00845$, and $0.00770$ per unit $Q$ at $Q=0.66$, $1.32$,
$2.65$, and $5.30$, respectively. The deficit per unit $Q$ varies linearly
with $Q$ at small amplitude and extrapolates to $0.00918$ as $Q\to0$. At
the root, the two parts add up to $2.51\%$, so that
$\gamma_Q/\gamma_N\simeq1-0.0095\,Q$ there. Here, $\gamma=s/T\simeq C_V/T$,
where $s$ and $C_V$ are the entropy and heat capacity per unit length at
fixed density. Moreover, at fixed $Q$, one
obtains $2/v^*=0.97503$, $0.97490$, and $0.97489$ for $K=3$, $4$, and
$5$--$8k_{\rm F}$, respectively. The entropy ratio varies by less than
$10^{-5}$ for $K\ge4k_{\rm F}$.

\subsection{Thermodynamic derivatives of the four-line bubble}
Stationarity removes the first-order contribution from the change in
$G$, giving
\begin{equation}
\Omega_Q-\Omega_N=QC-Q\,\Pi_4(q=0;T,\mu)+O(Q^2).
\label{eq:Omega1}
\end{equation}
Here $\Pi_4$ is evaluated with free internal lines. At $\nu=-E_3$, the
occupation factors satisfy
$\mathcal N_{bcd}[f(\nu)-f(\xi_1)]=\prod_{j=1}^4(1-f_j)-\prod_{j=1}^4f_j$.
Since $S$ already contains $\mathcal N_{bcd}$, the remaining frequency
sum reduces the bubble to
\begin{equation}
\Pi_4(q;T,\mu)=\frac1Q\int\frac{dk}{2\pi}\int d\nu\,S_q(k,\nu)\,
\frac{f(\nu)-f(\xi_k)}{\xi_k-\nu},
\label{eq:Pi4S}
\end{equation}
where $S_q$ is given by Eq.~(\ref{eq:S}) for the internal momenta summing
up to $q-k$. On a common energy grid at the $(Q,\mu)$ of the $T=0.025$
root, we obtain $\Pi_4=0.0073418$, $0.0073601$, $0.0073647$, and
$0.0073658$ at $T=0.05$, $0.025$, $0.0125$, and $0.00625E_{\rm F}$,
respectively, which follow a $T^2$ law with the coefficient $-0.00964$ to
$-0.00967$. A fitted $T^2\ln T$ term has a coefficient of $1.2\%$ of the
$T^2$ coefficient, so that the bubble follows a $T^2$ law within $6\%$ at
the computed temperatures. The
entropy shift $Q\partial_T\Pi_4$ gives the deficit $2.444$--$2.451\%$ at
fixed $\mu$, in comparison with approximately $2.44\%$ from the Landau
term linearized in $Q$ and normalized to the same fixed-$\mu$ free gas. In addition, the density shift
$Q\partial_\mu\Pi_4=-0.003768$ agrees with $-0.003782$ obtained from the
Dyson-resummed closure. With dressed lines, the root condition is
$\Pi_4=C=0.007012$, whereas free lines give $0.007360$.

\section{Counting integrals and numerical checks}\label{app:count}

\subsection{Momentum integration}
For the leading term in Eq.~(\ref{eq:J}), the one-dimensional four-hole
region is given by $|s|+\max(|u|,|v|)/\sqrt2<k_\mu$, with
$k_\mu=\sqrt{2m\mu}$. The four-particle region requires
$|k_i|>k_\mu$ and $|k_i|\le K$ for every leg. For fixed $(u,v)$ the
allowed domains are unions of intervals in $s$. For $m=1/2$, the
integral over $s$ of $(4s^2-E_\perp)^{-2}$ is elementary, where
$E_\perp=4\mu-(u^2+v^2)/(2m)$. The remaining two-dimensional integral
converges at the touching configurations, where all four momenta reach
the Fermi points.
With the same vertex in $d$ dimensions and $m=1/2$, the Jacobi variables
$s$, $u$, and $v$ are vectors and $E_4=4s^2+u^2+v^2-4\mu$. At fixed
$(u,v,\hat s)$, the occupation constraints divide $r=|s|$ into allowed
intervals, and the integral
$\int r^{d-1}dr\,(4r^2-E_\perp)^{-2}$ is elementary. The remaining variables
are integrated by the Gauss-Legendre quadrature after fixing $u$ along an
axis. In $d=1$, the quadrature with $n_{\rm rad}=3200$ nodes reproduces the
exact $s$ integration of Eq.~(\ref{eq:J}) within $10^{-4}$ at
$\kappa=0.6$ and $0.7k_{\rm F}$, whereas the coarser grid with
$n_{\rm rad}=400$ is not converged for $\kappa\ge0.6k_{\rm F}$. The ratio
at $K=4k_{\rm F}$ is summarized in Table~\ref{tab:dim}.

\begin{table*}[t]
\caption{Ratio $J_{4p}/J_{4h}$ in $d$ dimensions. Here, $K=4k_{\rm F}$ and
$\mu=1.00549E_{\rm F}$. The values in $d=3$ carry a quadrature uncertainty of
$0.5\%$, estimated from a finer grid at $\kappa=0.3$, $0.5$, and
$1.0k_{\rm F}$.}
\label{tab:dim}
\begin{ruledtabular}
\begin{tabular}{lcccccccc}
$\kappa/k_{\rm F}$ & $0.3$ & $0.4$ & $0.5$ & $0.6$ & $0.7$ & $0.8$ & $0.9$ & $1.0$\\
\hline
$d=1$ & $0.424$ & $0.364$ & $0.326$ & $0.305$ & $0.299$ & $0.302$ & $0.311$ & $0.324$\\
$d=2$ & $1.104$ & $1.217$ & $1.413$ & $1.705$ & $2.099$ & $2.583$ & $3.151$ & $3.787$\\
$d=3$ & $2.940$ & $4.379$ & $6.845$ & $10.89$ & $16.58$ & $24.71$ & $35.25$ & $48.28$\\
\end{tabular}
\end{ruledtabular}
\end{table*}

\begin{table*}[t]
\caption{Dependence of $J_{4p}/J_{4h}$ on the cutoff $K$ in $d=2$ and $3$.
Here, $\mu=1.00549E_{\rm F}$.}
\label{tab:cutoff}
\begin{ruledtabular}
\begin{tabular}{llccccc}
$d$ & $\kappa/k_{\rm F}$ & $K=2k_{\rm F}$ & $3k_{\rm F}$ & $4k_{\rm F}$ & $6k_{\rm F}$ & $8k_{\rm F}$\\
\hline
$2$ & $0.1$ & $0.969$ & $1.010$ & $1.020$ & $1.027$ & $1.029$\\
$2$ & $0.3$ & $0.840$ & $1.054$ & $1.105$ & $1.134$ & $1.143$\\
$2$ & $0.5$ & $0.809$ & $1.265$ & $1.414$ & $1.499$ & $1.523$\\
$2$ & $1.0$ & $1.347$ & $2.882$ & $3.777$ & $4.472$ & $4.687$\\
$3$ & $0.1$ & $1.389$ & $1.503$ & $1.546$ & $1.584$ & $1.602$\\
$3$ & $0.3$ & $1.874$ & $2.657$ & $2.940$ & $3.180$ & $3.288$\\
$3$ & $0.5$ & $2.889$ & $5.572$ & $6.845$ & $7.926$ & $8.404$\\
$3$ & $1.0$ & $8.307$ & $29.12$ & $48.28$ & $72.51$ & $84.10$\\
\end{tabular}
\end{ruledtabular}
\end{table*}

\subsection{Small-amplitude comparison}
Table~\ref{tab:closure} lists the bare-line ratios used in
Fig.~\ref{fig:test}(b).
\begin{table}[t]
\caption{$R(Q,T)$, the ratio of the thermally corrected bare-line Dyson
counting anomaly to the leading term of Eq.~(\ref{eq:J}). Here,
$\mu$ is that of the $T=0.025$ root, $\kappa=0.5k_{\rm F}$, and $Q_0=2.648$
is the amplitude of that root.}
\label{tab:closure}
\begin{ruledtabular}
\begin{tabular}{lccc}
& $T=0.025E_{\rm F}$ & $T=0.0125E_{\rm F}$ & $T=0.00625E_{\rm F}$\\
\hline
$Q=Q_0$ & $0.9584$ & $0.9614$ & $0.9624$\\
$Q=Q_0/2$ & $0.9763$ & $0.9795$ & $0.9806$\\
$Q=Q_0/4$ & $0.9854$ & $0.9888$ & $0.9898$\\
$Q\to0$ (linear) & $0.9944$ & $0.9978$ & $0.9989$\\
\end{tabular}
\end{ruledtabular}
\end{table}

\subsection{Spectral integral and momentum resolution}
\label{sec:directtest}
Using $1/E^2=\int_0^\infty dt\,t e^{-t|E|}$, define
$P_k(t)=\sum_{e_j>0}w_je^{-te_j}$ and
$H_k(t)=\sum_{e_j<0}w_je^{te_j}$. The factorization of $|\phi|^2$ into
the pairs $(ac)$ and $(bd)$, with identical inversion-symmetric spectra,
gives
\begin{equation}
\begin{aligned}
\mathcal B_T&=\frac{4Q}{(2\pi)^3}\int_0^\infty dt\,t\int ds\,
\bigl[F_P(s,t)^2-F_H(s,t)^2\bigr],\\
F_P(s,t)&=\int du\,e^{-u^2/(4\kappa^2)}
P_{s+u/\sqrt2}(t)P_{s-u/\sqrt2}(t),
\end{aligned}
\label{eq:laplace}
\end{equation}
and similarly for $F_H$. On the solver grid,
$s=(k_i+k_j)/2$ and $u=(k_i-k_j)/\sqrt2$, so $F$ is a weighted sum
along $i+j=$ const. The logarithmic $t$ grid has 201 points between
$10^{-3}$ and $10^5$. Free lines on the same momentum grid give
$-0.0120702$, compared with $-0.0120042$ from the continuous integral
(\ref{eq:J}), a $0.55\%$ offset. Configurations on energy scales
$v_{\rm F}\Delta k$ and below are sensitive to the momentum sampling.
The corresponding Laplace time is $(v_{\rm F}\Delta k)^{-1}$.
Here $\Delta k$ is the momentum spacing, distinct from the energy step
$h$. All values reported below use the full solver momentum grid.

An independent postprocessing check on the two separately solved
$T=0.025$ roots gives $\mathcal B_T/\mathcal A_T=1.0305$ and $1.0119$
at $N_k=2049$ and $4097$, respectively. The right-hand side changes
by $2.1\%$ between these grids. Table~\ref{tab:directquad} shows that
doubling the $t$ grid hardly changes the result. Integrating the tail
from $10^5$ to $10^7$ adds $6.5\times10^{-7}$ at $T=0.05$, less than
$0.007\%$ of $|\mathcal B_T|$, and at most $3.6\times10^{-11}$ for
the two $T=0.025$ roots. Thus the Laplace grid and tail are small in
these checks. Varying the spectral-peak estimator changes the roots'
counting anomaly by $0.7$--$0.8\%$. The measured momentum-grid difference
and peak-estimator spread quantify two separate sensitivities of this
comparison.

\begin{table}[t]
\caption{Laplace quadrature check using the same stored spectra and
momentum nodes. The two $T=0.025$ rows are independently solved roots.
All values are $\mathcal B_T/\mathcal A_T$.}
\label{tab:directquad}
\begin{ruledtabular}
\begin{tabular}{lccc}
$T/E_{\rm F}$ & $N_k$ & 201 $t$ nodes & 401 $t$ nodes\\
\hline
$0.025$ & $2049$ & $1.0305164$ & $1.0305164$\\
$0.025$ & $4097$ & $1.0119059$ & $1.0119059$\\
$0.05$ & $4097$ & $0.9871195$ & $0.9871177$\\
\end{tabular}
\end{ruledtabular}
\end{table}

\subsection{Dressing factors and pairing slopes}
Table~\ref{tab:dressing} compares the dressed and bare-line results at
the parameters of Table~\ref{tab:width}. Table~\ref{tab:pairdim} compares
the calculated Cooper logarithm with the Fermi-surface estimate in one,
two, and three dimensions.

\begin{table*}[t]
\caption{Dressing factors at the $(Q,\mu)$ of Table~\ref{tab:width}. Here,
$Z$ is that of the self-consistent states, and each ratio is the
self-consistent value divided by the bare-line value at the same width. The
pair eigenvalues are evaluated on the solver grid.}
\label{tab:dressing}
\begin{ruledtabular}
\begin{tabular}{lcccccccc}
$\kappa/k_{\rm F}$ & $Z$ & count ratio & $Z^3$ & shift ratio & $\lambda_{ab}$ free & $\lambda_{ab}$ dressed & $\lambda_{ab}$ ratio & $Z^2$\\
\hline
$0.30$ & $0.9775$ & $0.932$ & $0.934$ & $0.950$ & $0.509$ & $0.486$ & $0.955$ & $0.956$\\
$0.40$ & $0.9742$ & $0.917$ & $0.924$ & $0.946$ & $0.592$ & $0.561$ & $0.948$ & $0.949$\\
$0.45$ & $0.9724$ & $0.911$ & $0.919$ & $0.942$ & $0.636$ & $0.600$ & $0.944$ & $0.946$\\
$0.50$ & $0.9707$ & $0.906$ & $0.915$ & $0.939$ & $0.680$ & $0.639$ & $0.940$ & $0.942$\\
$0.55$ & $0.9687$ & $0.901$ & $0.909$ & $0.936$ & $0.723$ & $0.676$ & $0.936$ & $0.938$\\
$0.60$ & $0.9668$ & $0.898$ & $0.904$ & $0.934$ & $0.763$ & $0.711$ & $0.933$ & $0.935$\\
$0.70$ & $0.9613$ & $0.893$ & $0.888$ & $0.930$ & $0.835$ & $0.774$ & $0.927$ & $0.924$\\
\end{tabular}
\end{ruledtabular}
\end{table*}

\begin{table}[t]
\caption{Growth of $\lambda_{ab}$ per unit $\ln(1/T)$ at $Q_0=2.648$ and
$\mu=1.00549E_{\rm F}$. The first value in each entry is obtained from the
eigenvalue problem between $T=0.0125$ and $0.003125E_{\rm F}$, and the second
value is given by the formula, where $N(0)$ and $\eta$ are evaluated at
$k_\mu=\sqrt\mu$ with $v_\mu=2\sqrt\mu$. Here $N(0)$ is the density of
states of one component.}
\label{tab:pairdim}
\begin{ruledtabular}
\begin{tabular}{lccc}
$\kappa/k_{\rm F}$ & $d=1$ & $d=2$ & $d=3$\\
\hline
$0.3$ & $0.1275$ / $0.1296$ & $0.0326$ / $0.0328$ & $0.00735$ / $0.00737$\\
$0.5$ & $0.1454$ / $0.1464$ & $0.0601$ / $0.0601$ & $0.01776$ / $0.01780$\\
$1.0$ & $0.2073$ / $0.2073$ & $0.1022$ / $0.1023$ & $0.03232$ / $0.03248$\\
\end{tabular}
\end{ruledtabular}
\end{table}

\FloatBarrier

\section{Phase stiffness and coherence}
\label{sec:phase}
We estimate the phase stiffness and coherence scales relevant to the
homogeneous-source approximation. We use free internal lines at
$\kappa=0.5k_{\rm F}$ and the fixed $(Q,\mu)$ of the $T=0.025E_{\rm F}$ root.
A condensate twisted to the momentum $q$ costs $-Q[\Pi_4(q)-\Pi_4(0)]$ at
first order, so that the stiffness of the phase $\theta$ with
$q=\partial_x\theta$ is given by $\rho_\theta=Q|\partial_q^2\Pi_4|$. From
Eq.~(\ref{eq:Pi4S}), we obtain $|\partial_q^2\Pi_4|=0.0084$ ($q=0.04$) and
$0.0080$ ($q=0.08$) at $T=0.025E_{\rm F}$, and $0.0095$ and $0.0086$ at
$T=0.0125E_{\rm F}$. They give $\rho_\theta=0.022$--$0.025$, namely about
$15\%$ of $n/(16m)=0.159$, the Galilean value for a fully superfluid gas.
Classical phase
fluctuations in one dimension give
$\langle e^{i\theta(x)}e^{-i\theta(0)}\rangle=e^{-|x|/\xi_\phi}$ with
$\xi_\phi=2\rho_\theta/T=1.8$ and $4.0k_{\rm F}^{-1}$ at the two
temperatures. At first order with free internal lines, a condensate momentum $q$
shifts the light-cone kinematics by $\delta k\to\delta k-q$.
The classical correlation length therefore suggests a momentum scale
$\xi_\phi^{-1}=0.56$ and $0.25k_{\rm F}$, larger than the computed
Fermi-point shift. Its characteristic energy
$v_{\rm F}/\xi_\phi\simeq0.5$--$1.1E_{\rm F}$ is also well above $T$.
These coherence scales characterize the uniform-source approximation
used for the counting calculation. Its momentum-space Fermi-volume
interpretation assumes a homogeneous phase-coherent background, as
specified in Sec.~\ref{sec:luttinger-derivation}.
For tubes
coupled by a quartet Josephson energy $-2J_\perp\cos(\theta_i-\theta_j)$ per
length between $z$ neighbors, the chain mean-field theory gives
$T_{3D}=2\sqrt{zJ_\perp\rho_\theta}$. Within this estimate, phase coherence
at $T=0.025E_{\rm F}$ requires
$zJ_\perp\gtrsim T^2/(4\rho_\theta)\simeq0.007E_{\rm F}k_{\rm F}$.


\begin{thebibliography}{29}%
\makeatletter
\providecommand \@ifxundefined [1]{%
 \@ifx{#1\undefined}
}%
\providecommand \@ifnum [1]{%
 \ifnum #1\expandafter \@firstoftwo
 \else \expandafter \@secondoftwo
 \fi
}%
\providecommand \@ifx [1]{%
 \ifx #1\expandafter \@firstoftwo
 \else \expandafter \@secondoftwo
 \fi
}%
\providecommand \natexlab [1]{#1}%
\providecommand \enquote  [1]{``#1''}%
\providecommand \bibnamefont  [1]{#1}%
\providecommand \bibfnamefont [1]{#1}%
\providecommand \citenamefont [1]{#1}%
\providecommand \href@noop [0]{\@secondoftwo}%
\providecommand \href [0]{\begingroup \@sanitize@url \@href}%
\providecommand \@href[1]{\@@startlink{#1}\@@href}%
\providecommand \@@href[1]{\endgroup#1\@@endlink}%
\providecommand \@sanitize@url [0]{\catcode `\\12\catcode `\$12\catcode
  `\&12\catcode `\#12\catcode `\^12\catcode `\_12\catcode `\%12\relax}%
\providecommand \@@startlink[1]{}%
\providecommand \@@endlink[0]{}%
\providecommand \url  [0]{\begingroup\@sanitize@url \@url }%
\providecommand \@url [1]{\endgroup\@href {#1}{\urlprefix }}%
\providecommand \urlprefix  [0]{URL }%
\providecommand \Eprint [0]{\href }%
\providecommand \doibase [0]{https://doi.org/}%
\providecommand \selectlanguage [0]{\@gobble}%
\providecommand \bibinfo  [0]{\@secondoftwo}%
\providecommand \bibfield  [0]{\@secondoftwo}%
\providecommand \translation [1]{[#1]}%
\providecommand \BibitemOpen [0]{}%
\providecommand \bibitemStop [0]{}%
\providecommand \bibitemNoStop [0]{.\EOS\space}%
\providecommand \EOS [0]{\spacefactor3000\relax}%
\providecommand \BibitemShut  [1]{\csname bibitem#1\endcsname}%
\let\auto@bib@innerbib\@empty
\bibitem [{\citenamefont {Berg}\ \emph {et~al.}(2009)\citenamefont {Berg},
  \citenamefont {Fradkin},\ and\ \citenamefont
  {Kivelson}}]{BergFradkinKivelson2009}%
  \BibitemOpen
  \bibfield  {author} {\bibinfo {author} {\bibfnamefont {E.}~\bibnamefont
  {Berg}}, \bibinfo {author} {\bibfnamefont {E.}~\bibnamefont {Fradkin}},\ and\
  \bibinfo {author} {\bibfnamefont {S.~A.}\ \bibnamefont {Kivelson}},\
  }\bibfield  {title} {\bibinfo {title} {Charge-4e superconductivity from
  pair-density-wave order in certain high-temperature superconductors},\ }\href
  {https://doi.org/10.1038/nphys1389} {\bibfield  {journal} {\bibinfo
  {journal} {Nat. Phys.}\ }\textbf {\bibinfo {volume} {5}},\ \bibinfo {pages}
  {830} (\bibinfo {year} {2009})}\BibitemShut {NoStop}%
\bibitem [{\citenamefont {Ge}\ \emph {et~al.}(2024)\citenamefont {Ge},
  \citenamefont {Wang}, \citenamefont {Xing}, \citenamefont {Yin},
  \citenamefont {Wang}, \citenamefont {Shen}, \citenamefont {Lei},
  \citenamefont {Wang},\ and\ \citenamefont {Wang}}]{Ge2024}%
  \BibitemOpen
  \bibfield  {author} {\bibinfo {author} {\bibfnamefont {J.}~\bibnamefont
  {Ge}}, \bibinfo {author} {\bibfnamefont {P.}~\bibnamefont {Wang}}, \bibinfo
  {author} {\bibfnamefont {Y.}~\bibnamefont {Xing}}, \bibinfo {author}
  {\bibfnamefont {Q.}~\bibnamefont {Yin}}, \bibinfo {author} {\bibfnamefont
  {A.}~\bibnamefont {Wang}}, \bibinfo {author} {\bibfnamefont {J.}~\bibnamefont
  {Shen}}, \bibinfo {author} {\bibfnamefont {H.}~\bibnamefont {Lei}}, \bibinfo
  {author} {\bibfnamefont {Z.}~\bibnamefont {Wang}},\ and\ \bibinfo {author}
  {\bibfnamefont {J.}~\bibnamefont {Wang}},\ }\bibfield  {title} {\bibinfo
  {title} {Charge-4e and charge-6e flux quantization and higher charge
  superconductivity in kagome superconductor ring devices},\ }\href
  {https://doi.org/10.1103/PhysRevX.14.021025} {\bibfield  {journal} {\bibinfo
  {journal} {Phys. Rev. X}\ }\textbf {\bibinfo {volume} {14}},\ \bibinfo
  {pages} {021025} (\bibinfo {year} {2024})}\BibitemShut {NoStop}%
\bibitem [{\citenamefont {Guo}\ \emph {et~al.}(2022{\natexlab{a}})\citenamefont
  {Guo}, \citenamefont {Tajima},\ and\ \citenamefont {Liang}}]{GuoEH2022}%
  \BibitemOpen
  \bibfield  {author} {\bibinfo {author} {\bibfnamefont {Y.}~\bibnamefont
  {Guo}}, \bibinfo {author} {\bibfnamefont {H.}~\bibnamefont {Tajima}},\ and\
  \bibinfo {author} {\bibfnamefont {H.}~\bibnamefont {Liang}},\ }\bibfield
  {title} {\bibinfo {title} {Biexciton-like quartet condensates in an
  electron-hole liquid},\ }\href
  {https://doi.org/10.1103/PhysRevResearch.4.023152} {\bibfield  {journal}
  {\bibinfo  {journal} {Phys. Rev. Res.}\ }\textbf {\bibinfo {volume} {4}},\
  \bibinfo {pages} {023152} (\bibinfo {year} {2022}{\natexlab{a}})}\BibitemShut
  {NoStop}%
\bibitem [{\citenamefont {Wu}(2005)}]{Wu2005}%
  \BibitemOpen
  \bibfield  {author} {\bibinfo {author} {\bibfnamefont {C.}~\bibnamefont
  {Wu}},\ }\bibfield  {title} {\bibinfo {title} {Competing orders in
  one-dimensional spin-3/2 fermionic systems},\ }\href
  {https://doi.org/10.1103/PhysRevLett.95.266404} {\bibfield  {journal}
  {\bibinfo  {journal} {Phys. Rev. Lett.}\ }\textbf {\bibinfo {volume} {95}},\
  \bibinfo {pages} {266404} (\bibinfo {year} {2005})}\BibitemShut {NoStop}%
\bibitem [{\citenamefont {Capponi}\ \emph {et~al.}(2008)\citenamefont
  {Capponi}, \citenamefont {Roux}, \citenamefont {Lecheminant}, \citenamefont
  {Azaria}, \citenamefont {Boulat},\ and\ \citenamefont {White}}]{Capponi2008}%
  \BibitemOpen
  \bibfield  {author} {\bibinfo {author} {\bibfnamefont {S.}~\bibnamefont
  {Capponi}}, \bibinfo {author} {\bibfnamefont {G.}~\bibnamefont {Roux}},
  \bibinfo {author} {\bibfnamefont {P.}~\bibnamefont {Lecheminant}}, \bibinfo
  {author} {\bibfnamefont {P.}~\bibnamefont {Azaria}}, \bibinfo {author}
  {\bibfnamefont {E.}~\bibnamefont {Boulat}},\ and\ \bibinfo {author}
  {\bibfnamefont {S.~R.}\ \bibnamefont {White}},\ }\bibfield  {title} {\bibinfo
  {title} {Molecular superfluid phase in systems of one-dimensional
  multicomponent fermionic cold atoms},\ }\href
  {https://doi.org/10.1103/PhysRevA.77.013624} {\bibfield  {journal} {\bibinfo
  {journal} {Phys. Rev. A}\ }\textbf {\bibinfo {volume} {77}},\ \bibinfo
  {pages} {013624} (\bibinfo {year} {2008})}\BibitemShut {NoStop}%
\bibitem [{\citenamefont {Liu}\ \emph {et~al.}(2023)\citenamefont {Liu},
  \citenamefont {Wang},\ and\ \citenamefont {Cui}}]{Liu2023}%
  \BibitemOpen
  \bibfield  {author} {\bibinfo {author} {\bibfnamefont {R.}~\bibnamefont
  {Liu}}, \bibinfo {author} {\bibfnamefont {W.}~\bibnamefont {Wang}},\ and\
  \bibinfo {author} {\bibfnamefont {X.}~\bibnamefont {Cui}},\ }\bibfield
  {title} {\bibinfo {title} {Quartet superfluid in two-dimensional
  mass-imbalanced {Fermi} mixtures},\ }\href
  {https://doi.org/10.1103/PhysRevLett.131.193401} {\bibfield  {journal}
  {\bibinfo  {journal} {Phys. Rev. Lett.}\ }\textbf {\bibinfo {volume} {131}},\
  \bibinfo {pages} {193401} (\bibinfo {year} {2023})}\BibitemShut {NoStop}%
\bibitem [{\citenamefont {R{\"o}pke}\ \emph {et~al.}(1998)\citenamefont
  {R{\"o}pke}, \citenamefont {Schnell}, \citenamefont {Schuck},\ and\
  \citenamefont {Nozi{\`e}res}}]{Ropke1998}%
  \BibitemOpen
  \bibfield  {author} {\bibinfo {author} {\bibfnamefont {G.}~\bibnamefont
  {R{\"o}pke}}, \bibinfo {author} {\bibfnamefont {A.}~\bibnamefont {Schnell}},
  \bibinfo {author} {\bibfnamefont {P.}~\bibnamefont {Schuck}},\ and\ \bibinfo
  {author} {\bibfnamefont {P.}~\bibnamefont {Nozi{\`e}res}},\ }\bibfield
  {title} {\bibinfo {title} {Four-particle condensate in strongly coupled
  fermion systems},\ }\href {https://doi.org/10.1103/PhysRevLett.80.3177}
  {\bibfield  {journal} {\bibinfo  {journal} {Phys. Rev. Lett.}\ }\textbf
  {\bibinfo {volume} {80}},\ \bibinfo {pages} {3177} (\bibinfo {year}
  {1998})}\BibitemShut {NoStop}%
\bibitem [{\citenamefont {Tohsaki}\ \emph {et~al.}(2001)\citenamefont
  {Tohsaki}, \citenamefont {Horiuchi}, \citenamefont {Schuck},\ and\
  \citenamefont {R{\"o}pke}}]{Tohsaki2001}%
  \BibitemOpen
  \bibfield  {author} {\bibinfo {author} {\bibfnamefont {A.}~\bibnamefont
  {Tohsaki}}, \bibinfo {author} {\bibfnamefont {H.}~\bibnamefont {Horiuchi}},
  \bibinfo {author} {\bibfnamefont {P.}~\bibnamefont {Schuck}},\ and\ \bibinfo
  {author} {\bibfnamefont {G.}~\bibnamefont {R{\"o}pke}},\ }\bibfield  {title}
  {\bibinfo {title} {Alpha cluster condensation in $^{12}${C} and $^{16}${O}},\
  }\href {https://doi.org/10.1103/PhysRevLett.87.192501} {\bibfield  {journal}
  {\bibinfo  {journal} {Phys. Rev. Lett.}\ }\textbf {\bibinfo {volume} {87}},\
  \bibinfo {pages} {192501} (\bibinfo {year} {2001})}\BibitemShut {NoStop}%
\bibitem [{\citenamefont {Sogo}\ \emph {et~al.}(2010)\citenamefont {Sogo},
  \citenamefont {R{\"o}pke},\ and\ \citenamefont {Schuck}}]{Sogo2010}%
  \BibitemOpen
  \bibfield  {author} {\bibinfo {author} {\bibfnamefont {T.}~\bibnamefont
  {Sogo}}, \bibinfo {author} {\bibfnamefont {G.}~\bibnamefont {R{\"o}pke}},\
  and\ \bibinfo {author} {\bibfnamefont {P.}~\bibnamefont {Schuck}},\
  }\bibfield  {title} {\bibinfo {title} {Many-body approach for quartet
  condensation in strong coupling},\ }\href
  {https://doi.org/10.1103/PhysRevC.81.064310} {\bibfield  {journal} {\bibinfo
  {journal} {Phys. Rev. C}\ }\textbf {\bibinfo {volume} {81}},\ \bibinfo
  {pages} {064310} (\bibinfo {year} {2010})}\BibitemShut {NoStop}%
\bibitem [{\citenamefont {Guo}\ \emph {et~al.}(2022{\natexlab{b}})\citenamefont
  {Guo}, \citenamefont {Tajima},\ and\ \citenamefont {Liang}}]{GuoNM2022}%
  \BibitemOpen
  \bibfield  {author} {\bibinfo {author} {\bibfnamefont {Y.}~\bibnamefont
  {Guo}}, \bibinfo {author} {\bibfnamefont {H.}~\bibnamefont {Tajima}},\ and\
  \bibinfo {author} {\bibfnamefont {H.}~\bibnamefont {Liang}},\ }\bibfield
  {title} {\bibinfo {title} {Cooper quartet correlations in infinite symmetric
  nuclear matter},\ }\href {https://doi.org/10.1103/PhysRevC.105.024317}
  {\bibfield  {journal} {\bibinfo  {journal} {Phys. Rev. C}\ }\textbf {\bibinfo
  {volume} {105}},\ \bibinfo {pages} {024317} (\bibinfo {year}
  {2022}{\natexlab{b}})}\BibitemShut {NoStop}%
\bibitem [{\citenamefont {Guo}\ \emph {et~al.}(2025)\citenamefont {Guo},
  \citenamefont {Naito}, \citenamefont {Tajima},\ and\ \citenamefont
  {Liang}}]{GuoNZ2025}%
  \BibitemOpen
  \bibfield  {author} {\bibinfo {author} {\bibfnamefont {Y.}~\bibnamefont
  {Guo}}, \bibinfo {author} {\bibfnamefont {T.}~\bibnamefont {Naito}}, \bibinfo
  {author} {\bibfnamefont {H.}~\bibnamefont {Tajima}},\ and\ \bibinfo {author}
  {\bibfnamefont {H.}~\bibnamefont {Liang}},\ }\bibfield  {title} {\bibinfo
  {title} {Quartet correlations near the surface of {$N=Z$} nuclei},\ }\href
  {https://doi.org/10.1103/4rqf-5kfx} {\bibfield  {journal} {\bibinfo
  {journal} {Phys. Rev. C}\ }\textbf {\bibinfo {volume} {112}},\ \bibinfo
  {pages} {024310} (\bibinfo {year} {2025})}\BibitemShut {NoStop}%
\bibitem [{\citenamefont {Jin}\ \emph {et~al.}(2026)\citenamefont {Jin},
  \citenamefont {Li},\ and\ \citenamefont {Zhou}}]{Jin2026}%
  \BibitemOpen
  \bibfield  {author} {\bibinfo {author} {\bibfnamefont {J.-T.}\ \bibnamefont
  {Jin}}, \bibinfo {author} {\bibfnamefont {P.}~\bibnamefont {Li}},\ and\
  \bibinfo {author} {\bibfnamefont {Y.}~\bibnamefont {Zhou}},\ }\href@noop {}
  {\bibinfo {title} {Charge-4e superconducting ground state without pair
  condensation: Exact quartet dynamics, rigorous order, and a microscopic
  route}} (\bibinfo {year} {2026}),\ \bibinfo {note}
  {\href{https://arxiv.org/abs/2609.24445}{arXiv:2609.24445}}\BibitemShut
  {NoStop}%
\bibitem [{\citenamefont {Xu}\ and\ \citenamefont {Wu}(2026)}]{XuWu2026}%
  \BibitemOpen
  \bibfield  {author} {\bibinfo {author} {\bibfnamefont {Q.-R.}\ \bibnamefont
  {Xu}}\ and\ \bibinfo {author} {\bibfnamefont {C.}~\bibnamefont {Wu}},\
  }\href@noop {} {\bibinfo {title} {Effective theory of fermion quartet
  condensation}} (\bibinfo {year} {2026}),\ \bibinfo {note}
  {\href{https://arxiv.org/abs/2605.29338}{arXiv:2605.29338}}\BibitemShut
  {NoStop}%
\bibitem [{\citenamefont {Jiang}\ \emph {et~al.}(2017)\citenamefont {Jiang},
  \citenamefont {Li}, \citenamefont {Kivelson},\ and\ \citenamefont
  {Yao}}]{Jiang2017}%
  \BibitemOpen
  \bibfield  {author} {\bibinfo {author} {\bibfnamefont {Y.-F.}\ \bibnamefont
  {Jiang}}, \bibinfo {author} {\bibfnamefont {Z.-X.}\ \bibnamefont {Li}},
  \bibinfo {author} {\bibfnamefont {S.~A.}\ \bibnamefont {Kivelson}},\ and\
  \bibinfo {author} {\bibfnamefont {H.}~\bibnamefont {Yao}},\ }\bibfield
  {title} {\bibinfo {title} {Charge-4e superconductors: A {Majorana} quantum
  {Monte Carlo} study},\ }\href {https://doi.org/10.1103/PhysRevB.95.241103}
  {\bibfield  {journal} {\bibinfo  {journal} {Phys. Rev. B}\ }\textbf {\bibinfo
  {volume} {95}},\ \bibinfo {pages} {241103(R)} (\bibinfo {year}
  {2017})}\BibitemShut {NoStop}%
\bibitem [{\citenamefont {Gnezdilov}\ and\ \citenamefont
  {Wang}(2022)}]{GnezdilovWang2022}%
  \BibitemOpen
  \bibfield  {author} {\bibinfo {author} {\bibfnamefont {N.~V.}\ \bibnamefont
  {Gnezdilov}}\ and\ \bibinfo {author} {\bibfnamefont {Y.}~\bibnamefont
  {Wang}},\ }\bibfield  {title} {\bibinfo {title} {Solvable model for a
  charge-4e superconductor},\ }\href
  {https://doi.org/10.1103/PhysRevB.106.094508} {\bibfield  {journal} {\bibinfo
   {journal} {Phys. Rev. B}\ }\textbf {\bibinfo {volume} {106}},\ \bibinfo
  {pages} {094508} (\bibinfo {year} {2022})}\BibitemShut {NoStop}%
\bibitem [{\citenamefont {Guo}\ \emph {et~al.}(2026)\citenamefont {Guo},
  \citenamefont {Tajima},\ and\ \citenamefont {Liang}}]{GuoTL2026}%
  \BibitemOpen
  \bibfield  {author} {\bibinfo {author} {\bibfnamefont {Y.}~\bibnamefont
  {Guo}}, \bibinfo {author} {\bibfnamefont {H.}~\bibnamefont {Tajima}},\ and\
  \bibinfo {author} {\bibfnamefont {H.}~\bibnamefont {Liang}},\ }\bibfield
  {title} {\bibinfo {title} {Comparative study of a quartet superfluid state:
  Quartet {Bardeen-Cooper-Schrieffer} theory and generalized {Nambu-Gor'kov}
  formalism},\ }\href {https://doi.org/10.1103/snvw-f54b} {\bibfield  {journal}
  {\bibinfo  {journal} {Phys. Rev. C}\ }\textbf {\bibinfo {volume} {114}},\
  \bibinfo {pages} {034304} (\bibinfo {year} {2026})}\BibitemShut {NoStop}%
\bibitem [{\citenamefont {Wan}\ \emph {et~al.}(2026)\citenamefont {Wan},
  \citenamefont {Jiang}, \citenamefont {Zou}, \citenamefont {Zhang},\ and\
  \citenamefont {Jian}}]{Wan2026}%
  \BibitemOpen
  \bibfield  {author} {\bibinfo {author} {\bibfnamefont {Z.-Q.}\ \bibnamefont
  {Wan}}, \bibinfo {author} {\bibfnamefont {H.}~\bibnamefont {Jiang}}, \bibinfo
  {author} {\bibfnamefont {X.}~\bibnamefont {Zou}}, \bibinfo {author}
  {\bibfnamefont {S.}~\bibnamefont {Zhang}},\ and\ \bibinfo {author}
  {\bibfnamefont {S.-K.}\ \bibnamefont {Jian}},\ }\href@noop {} {\bibinfo
  {title} {Quantum charge-4e superconductivity and deconfined pseudocriticality
  in the attractive {SU(4)} {Hubbard} model}} (\bibinfo {year} {2026}),\
  \bibinfo {note}
  {\href{https://arxiv.org/abs/2604.14289}{arXiv:2604.14289}}\BibitemShut
  {NoStop}%
\bibitem [{\citenamefont {Luttinger}(1960)}]{Luttinger1960}%
  \BibitemOpen
  \bibfield  {author} {\bibinfo {author} {\bibfnamefont {J.~M.}\ \bibnamefont
  {Luttinger}},\ }\bibfield  {title} {\bibinfo {title} {Fermi surface and some
  simple equilibrium properties of a system of interacting fermions},\ }\href
  {https://doi.org/10.1103/PhysRev.119.1153} {\bibfield  {journal} {\bibinfo
  {journal} {Phys. Rev.}\ }\textbf {\bibinfo {volume} {119}},\ \bibinfo {pages}
  {1153} (\bibinfo {year} {1960})}\BibitemShut {NoStop}%
\bibitem [{\citenamefont {Luttinger}\ and\ \citenamefont
  {Ward}(1960)}]{LuttingerWard1960}%
  \BibitemOpen
  \bibfield  {author} {\bibinfo {author} {\bibfnamefont {J.~M.}\ \bibnamefont
  {Luttinger}}\ and\ \bibinfo {author} {\bibfnamefont {J.~C.}\ \bibnamefont
  {Ward}},\ }\bibfield  {title} {\bibinfo {title} {Ground-state energy of a
  many-fermion system. {II}},\ }\href
  {https://doi.org/10.1103/PhysRev.118.1417} {\bibfield  {journal} {\bibinfo
  {journal} {Phys. Rev.}\ }\textbf {\bibinfo {volume} {118}},\ \bibinfo {pages}
  {1417} (\bibinfo {year} {1960})}\BibitemShut {NoStop}%
\bibitem [{\citenamefont {Oshikawa}(2000)}]{Oshikawa2000}%
  \BibitemOpen
  \bibfield  {author} {\bibinfo {author} {\bibfnamefont {M.}~\bibnamefont
  {Oshikawa}},\ }\bibfield  {title} {\bibinfo {title} {Topological approach to
  {Luttinger}'s theorem and the {Fermi} surface of a {Kondo} lattice},\ }\href
  {https://doi.org/10.1103/PhysRevLett.84.3370} {\bibfield  {journal} {\bibinfo
   {journal} {Phys. Rev. Lett.}\ }\textbf {\bibinfo {volume} {84}},\ \bibinfo
  {pages} {3370} (\bibinfo {year} {2000})}\BibitemShut {NoStop}%
\bibitem [{\citenamefont {Baym}(1962)}]{Baym1962}%
  \BibitemOpen
  \bibfield  {author} {\bibinfo {author} {\bibfnamefont {G.}~\bibnamefont
  {Baym}},\ }\bibfield  {title} {\bibinfo {title} {Self-consistent
  approximations in many-body systems},\ }\href
  {https://doi.org/10.1103/PhysRev.127.1391} {\bibfield  {journal} {\bibinfo
  {journal} {Phys. Rev.}\ }\textbf {\bibinfo {volume} {127}},\ \bibinfo {pages}
  {1391} (\bibinfo {year} {1962})}\BibitemShut {NoStop}%
\bibitem [{\citenamefont {La~Nave}\ \emph {et~al.}(2025)\citenamefont
  {La~Nave}, \citenamefont {Zhao},\ and\ \citenamefont
  {Phillips}}]{LaNave2025}%
  \BibitemOpen
  \bibfield  {author} {\bibinfo {author} {\bibfnamefont {G.}~\bibnamefont
  {La~Nave}}, \bibinfo {author} {\bibfnamefont {J.}~\bibnamefont {Zhao}},\ and\
  \bibinfo {author} {\bibfnamefont {P.~W.}\ \bibnamefont {Phillips}},\
  }\bibfield  {title} {\bibinfo {title} {{Luttinger} count is the homotopy not
  the physical charge: Generalized anomalies characterize non-{Fermi}
  liquids},\ }\href {https://doi.org/10.1103/lfxt-fspg} {\bibfield  {journal}
  {\bibinfo  {journal} {Phys. Rev. Lett.}\ }\textbf {\bibinfo {volume} {135}},\
  \bibinfo {pages} {246501} (\bibinfo {year} {2025})}\BibitemShut {NoStop}%
\bibitem [{\citenamefont {Sachdev}\ and\ \citenamefont
  {Yang}(2006)}]{SachdevYang2006}%
  \BibitemOpen
  \bibfield  {author} {\bibinfo {author} {\bibfnamefont {S.}~\bibnamefont
  {Sachdev}}\ and\ \bibinfo {author} {\bibfnamefont {K.}~\bibnamefont {Yang}},\
  }\bibfield  {title} {\bibinfo {title} {Fermi surfaces and {Luttinger}'s
  theorem in paired fermion systems},\ }\href
  {https://doi.org/10.1103/PhysRevB.73.174504} {\bibfield  {journal} {\bibinfo
  {journal} {Phys. Rev. B}\ }\textbf {\bibinfo {volume} {73}},\ \bibinfo
  {pages} {174504} (\bibinfo {year} {2006})}\BibitemShut {NoStop}%
\bibitem [{\citenamefont {Powell}\ \emph {et~al.}(2005)\citenamefont {Powell},
  \citenamefont {Sachdev},\ and\ \citenamefont {B{\"u}chler}}]{Powell2005}%
  \BibitemOpen
  \bibfield  {author} {\bibinfo {author} {\bibfnamefont {S.}~\bibnamefont
  {Powell}}, \bibinfo {author} {\bibfnamefont {S.}~\bibnamefont {Sachdev}},\
  and\ \bibinfo {author} {\bibfnamefont {H.~P.}\ \bibnamefont {B{\"u}chler}},\
  }\bibfield  {title} {\bibinfo {title} {Depletion of the {Bose-Einstein}
  condensate in {Bose-Fermi} mixtures},\ }\href
  {https://doi.org/10.1103/PhysRevB.72.024534} {\bibfield  {journal} {\bibinfo
  {journal} {Phys. Rev. B}\ }\textbf {\bibinfo {volume} {72}},\ \bibinfo
  {pages} {024534} (\bibinfo {year} {2005})}\BibitemShut {NoStop}%
\bibitem [{\citenamefont {Carneiro}\ and\ \citenamefont
  {Pethick}(1975)}]{CarneiroPethick1975}%
  \BibitemOpen
  \bibfield  {author} {\bibinfo {author} {\bibfnamefont {G.~M.}\ \bibnamefont
  {Carneiro}}\ and\ \bibinfo {author} {\bibfnamefont {C.~J.}\ \bibnamefont
  {Pethick}},\ }\bibfield  {title} {\bibinfo {title} {Specific heat of a normal
  {Fermi} liquid. {II}. {Microscopic} approach},\ }\href
  {https://doi.org/10.1103/PhysRevB.11.1106} {\bibfield  {journal} {\bibinfo
  {journal} {Phys. Rev. B}\ }\textbf {\bibinfo {volume} {11}},\ \bibinfo
  {pages} {1106} (\bibinfo {year} {1975})}\BibitemShut {NoStop}%
\bibitem [{\citenamefont {Hohenberg}(1967)}]{Hohenberg1967}%
  \BibitemOpen
  \bibfield  {author} {\bibinfo {author} {\bibfnamefont {P.~C.}\ \bibnamefont
  {Hohenberg}},\ }\bibfield  {title} {\bibinfo {title} {Existence of long-range
  order in one and two dimensions},\ }\href
  {https://doi.org/10.1103/PhysRev.158.383} {\bibfield  {journal} {\bibinfo
  {journal} {Phys. Rev.}\ }\textbf {\bibinfo {volume} {158}},\ \bibinfo {pages}
  {383} (\bibinfo {year} {1967})}\BibitemShut {NoStop}%
\bibitem [{\citenamefont {Lecheminant}\ \emph {et~al.}(2005)\citenamefont
  {Lecheminant}, \citenamefont {Boulat},\ and\ \citenamefont
  {Azaria}}]{Lecheminant2005}%
  \BibitemOpen
  \bibfield  {author} {\bibinfo {author} {\bibfnamefont {P.}~\bibnamefont
  {Lecheminant}}, \bibinfo {author} {\bibfnamefont {E.}~\bibnamefont
  {Boulat}},\ and\ \bibinfo {author} {\bibfnamefont {P.}~\bibnamefont
  {Azaria}},\ }\bibfield  {title} {\bibinfo {title} {Confinement and
  superfluidity in one-dimensional degenerate fermionic cold atoms},\ }\href
  {https://doi.org/10.1103/PhysRevLett.95.240402} {\bibfield  {journal}
  {\bibinfo  {journal} {Phys. Rev. Lett.}\ }\textbf {\bibinfo {volume} {95}},\
  \bibinfo {pages} {240402} (\bibinfo {year} {2005})}\BibitemShut {NoStop}%
\bibitem [{\citenamefont {Cazalilla}\ and\ \citenamefont
  {Rey}(2014)}]{CazalillaRey2014}%
  \BibitemOpen
  \bibfield  {author} {\bibinfo {author} {\bibfnamefont {M.~A.}\ \bibnamefont
  {Cazalilla}}\ and\ \bibinfo {author} {\bibfnamefont {A.~M.}\ \bibnamefont
  {Rey}},\ }\bibfield  {title} {\bibinfo {title} {Ultracold {Fermi} gases with
  emergent {SU(N)} symmetry},\ }\href
  {https://doi.org/10.1088/0034-4885/77/12/124401} {\bibfield  {journal}
  {\bibinfo  {journal} {Rep. Prog. Phys.}\ }\textbf {\bibinfo {volume} {77}},\
  \bibinfo {pages} {124401} (\bibinfo {year} {2014})}\BibitemShut {NoStop}%
\bibitem [{\citenamefont {Stewart}\ \emph {et~al.}(2008)\citenamefont
  {Stewart}, \citenamefont {Gaebler},\ and\ \citenamefont {Jin}}]{Stewart2008}%
  \BibitemOpen
  \bibfield  {author} {\bibinfo {author} {\bibfnamefont {J.~T.}\ \bibnamefont
  {Stewart}}, \bibinfo {author} {\bibfnamefont {J.~P.}\ \bibnamefont
  {Gaebler}},\ and\ \bibinfo {author} {\bibfnamefont {D.~S.}\ \bibnamefont
  {Jin}},\ }\bibfield  {title} {\bibinfo {title} {Using photoemission
  spectroscopy to probe a strongly interacting {Fermi} gas},\ }\href
  {https://doi.org/10.1038/nature07172} {\bibfield  {journal} {\bibinfo
  {journal} {Nature}\ }\textbf {\bibinfo {volume} {454}},\ \bibinfo {pages}
  {744} (\bibinfo {year} {2008})}\BibitemShut {NoStop}%
\end{thebibliography}

%

\end{document}